\documentclass[useAMS,usenatbib,usegraphicx]{mn2e}
\def\c{{\sl Chandra}}

\def\zp{$z_{\rm phot}$}

\def\p{$\pm$}

\def\j{{\sl J}}
\def\h{{\sl H}}
\def\k{{\sl K}}

\def\ltsim{\mathrel{\hbox{\rlap{\hbox{\lower4pt\hbox{$\sim$}}}\hbox{$<$}}}}
\def\gtsim{\mathrel{\hbox{\rlap{\hbox{\lower4pt\hbox{$\sim$}}}\hbox{$>$}}}}

\def\ergpspsqcm{ erg s$^{-1}$ cm$^{-2}$}
\def\ergps{ erg s$^{-1}$}

\def\xmm{{\sl XMM-Newton}}
\def\micron{${\rm{\mu}m}$}

\def\int{{\sl INT}}

\def\zp{$z_{\rm{phot}}$}

\def\ha{H$\alpha$}
\def\hb{H$\beta$}
\def\hg{H$\gamma$}
\def\paa{Pa$\alpha$}
\def\pab{Pa$\beta$}
\def\oiii{[OIII]}
\def\oii{[OII]}
\def\nii{[NII]}
\def\sii{[SII]}
\def\siii{[SIII]}
\def\l{$\lambda$}
\def\oi{[OI]}

\def\gtsim{\mathrel{\hbox{\rlap{\hbox{\lower4pt\hbox{$\sim$}}}\hbox{$>$}}}}
\def\lesssim{\mathrel{\hbox{\rlap{\hbox{\lower4pt\hbox{$\sim$}}}\hbox{$<$}}}}

\def\c{{\sl Chandra}}

\def\zp{$z_{\rm phot}$}

\def\p{$\pm$}

\title{VLT near-infrared spectra of hard serendipitous \c\ sources}

\author[P. Gandhi, C.S. Crawford and A.C. Fabian]
{\parbox[]{6.in} { P. Gandhi$^{\ddag}$, C.S. Crawford and A.C. Fabian  \\
\footnotesize
Institute of Astronomy, Madingley Road, Cambridge CB3 0HA, UK \\
$^{\ddag}$pg@ast.cam.ac.uk\\}}
\date{
      Received }
\pubyear{}

\begin{document}

\maketitle

\begin{abstract}
We present near-infrared long-slit spectra of eight optically-dim
X-ray sources obtained with ISAAC on the Very Large Telescope. Six of
the sources have hard X-ray emission with a significant fraction of
the counts emerging above 2 keV. All were discovered serendipitously
in the fields of three nearby galaxy clusters observed with \c, and
identified through near-infrared imaging. The X-ray fluxes lie close
to the break in the source counts. Two of the sources show narrow
emission lines, and a third has a broad line. One of the narrow
line-emitting sources has a clear redshift identification at $z=2.18$,
while the other has a tentative determination based on the highest
redshift detection of He I \l10830 at $z=1.26$. The remainder have
featureless spectra to deep limiting equivalent widths of
$\sim$20--60\AA\ and line flux $\sim 5\times 10^{-17}$ \ergpspsqcm\ in
the \k-band. High-quality \j, \h\ and \k s--band images of the sources
were combined with archival optical detections or limits to estimate a
photometric redshift for six. Two sources show complex double
morphology. The hard sources have spectral count ratios consistent
with heavily obscured AGN, while the host galaxy emits much of the
optical and near-infrared flux.
The most likely explanation for the featureless continua is that the line
photons are being scattered or destroyed by optically-thick gas and
associated dust with large covering fractions.

\end{abstract}

\section{Introduction}
The \c\ observatory\footnotetext[1]{Based on observations made with ESO Telescopes at the Paranal Observatories under programme 67.B-0188}\footnotetext[2]{Based on observations made with the Chandra X-ray Observatory} has largely resolved the 2--10 keV X-ray
background (XRB) within two years of its launch, after almost four
decades of scientific effort (Giacconi et al. 1962; Mushotzky et
al. 2000; Brandt et al. 2001; Giacconi et
al. 2002). Follow-up work within the past year has revealed that this
\c\ source population can be broadly split into type 1 active galactic nuclei (AGN)
, narrow emission-line AGN, optically-normal galaxies with no
sign of activity other than in X-rays and optically-faint sources
which are difficult to identify (Barger et al. 2001; Alexander et al. 2001; Rosati et al. 2001; Willott et al. 2001). {\sl XMM-Newton},
with its higher effective area at 10 keV, has also begun to deliver
results (Hasinger et al. 2001) 
which confirm the essential \c\ findings (e.g.,
Lehmann et al. 2001a; Mainieri et al. 2002).

The individual X-ray spectra of the hard sources are flat enough to
account for the XRB spectral slope of 0.4 (Marshall et al. 1980;
Gruber et al. 1999), and their integrated flux contributes 
$\sim 70-100$ percent of the XRB intensity. The ambiguity in the
absolute intensity, still to be resolved, is caused by 
cross-calibration mismatches between various X-ray missions and/or
cosmic variance of the sources themselves (eg, Barcons et al. 2000;
Cowie et al. 2002). Essentially, these observations agree with 
models which synthesize the XRB through the integrated
emission of populations of obscured AGN that are spread over redshift and have a range of intrinsic absorbing column densities of gas (Setti \&
Woltjer 1989; Gilli, Salvati \& Hasinger 2001; Wilman, Fabian \&
Nulsen 2000; Comastri et al. 1995). Little evidence had been found in pre-\c\ surveys for the existence of highly absorbed, intrinsically powerful objects (e.g., Halpern, Turner \& George 1999).


\subsection{Obscured \lq type 2\rq\ sources}

What would be the observable characteristics of such obscured objects? In the framework of the standard torus model for AGN (Antonucci \& Miller 1985), luminous yet narrow emission lines generated at parsec-scale distances from the nucleus should be visible in the optical band. Direct optical nuclear radiation would be scattered out of our line-of-sight, making optically-faint sources good type 2 AGN (and powerful type 2 QSO) candidates. In X-rays, the proportion of sources that are observed through transmitted radiation would depend on the range of inclination angles that are Compton-thin.

While the current \c\ and \xmm\ surveys have observed such sources, they have also found a substantial number of objects that possess few, or no, emission lines in their optical spectra. Some sources show absorption features characteristic of early-type galaxy spectra, while others are difficult to identify. For examples of these and other type 2 QSOs, refer to Stern et al. (2002), Norman et al. (2002), Gandhi, Fabian \& Crawford (2002), Hasinger (2002) and Nakanishi et al. (2001). In the $\sim$0.5--10 keV X-ray band, many sources show transmitted flux being reprocessed by absorbing matter of column densities $10^{22}-10^{23}$ cm$^{-2}$. Thus, large scale and/or large covering factor absorption is necessary to obscure all optical emission, but this should not be ompton-thick. This raises the question of whether the sources that will emerge in future 10--40 keV surveys will be an entirely new population of even harder X-ray sources, or whether there will be large overlap with the current population.

\subsection{Our previous work}

We have been studying hard X-ray sources found serendipitously in the fields of massive clusters of galaxies with \c\ (Crawford et al. 2001; 2002; Gandhi et al. 2002). The flux regime that we sample is $10^{-15}\sim 10^{-14}$ \ergpspsqcm. This is an important regime, as the bulk of the XRB intensity is generated in populations with these fluxes (e.g., Cowie et al. 2002). Our work complements surveys such as ChaMP (Wilkes et al. 2001) and \xmm\ serendipitous surveys (Watson et al. 2001; Barcons et al. 2001; Baldi et al. 2002). 

By targeting the hard, optically-dim sources with multi-band imaging leading to photometric redshifts as well as optical spectroscopy, we have been able to discover both broad and narrow-line AGN and find that these lie at a large range of redshift ($0.2\ltsim z\ltsim 4$). The broad-line sources have equivalent-widths and line-fluxes similar to those identified by previous missions such as ROSAT (Lehmann et al. 2001b) and are predominantly associated with the X-ray soft sources. The hard sources have X-ray spectra that suggest high but Compton-thin absorption and possess narrow emission lines (Seyfert 2s). We also find one source which does not fit this pattern: with high X-ray absorption (N$_{\rm H}> 10^{22}$ cm$^{-2}$) but a broad strong Mg II \l2797\AA\ line, this source probably has a dust:gas ratio different to the Galactic value (Crawford et al. 2002).

The gravitational magnification of the cluster potential well enables us to {\sl study}, rather than just detect, some sources which lie within $\approx$ 1 arcmin of the cluster centre. Two such sources magnified by factors of 2 and 8 by the cluster Abell 2390 were found to be powerful obscured (N$_{\rm H}>10^{23}$ cm$^{-2}$) sources, even after de-magnification. The first source (A18 in Crawford et al. 2002) has an intrinsic X-ray luminosity L$_{\rm 2-10\ keV}> 10^{45}$ erg s$^{-1}$, while {\sl ISOCAM} 6.7 and 15-\micron\ detections are used to infer an absorbed big blue bump luminosity L$_{\rm UV}>10^{45}$ erg s$^{-1}$ for the second source (A15).\\

\begin{figure*}
  \begin{center}
\includegraphics[angle=90,width=15cm]{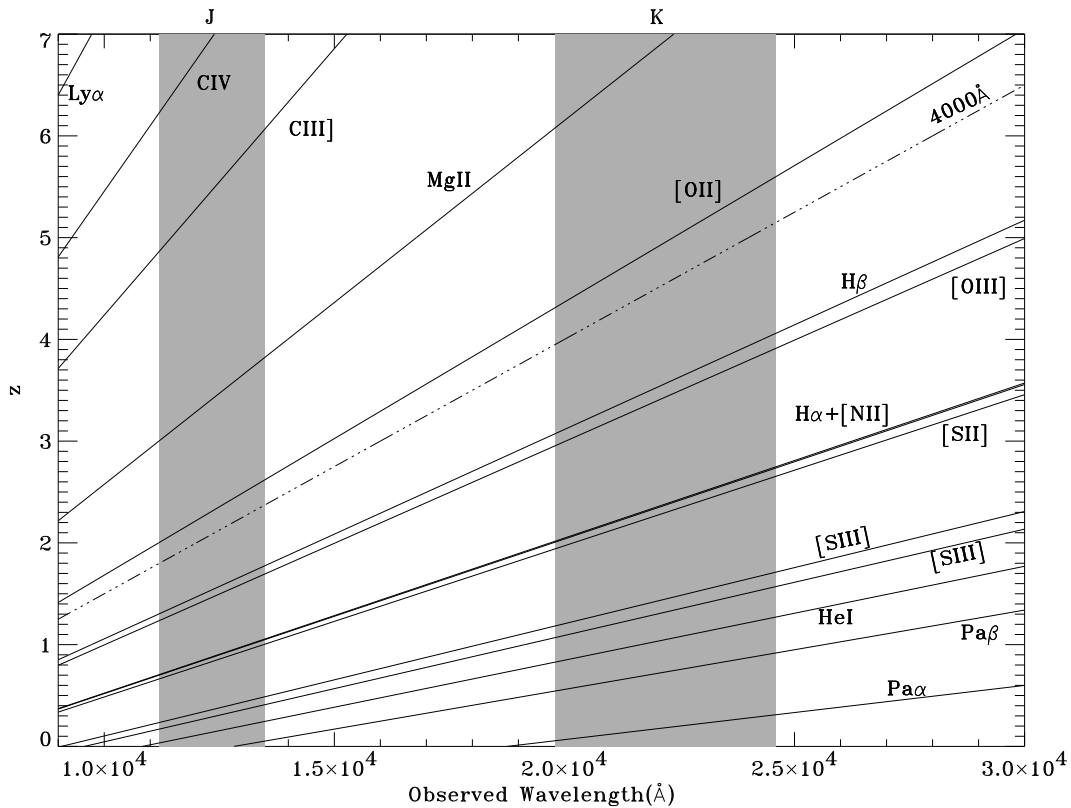}
  \caption{The wavelength of observation of typically strong Seyfert / QSO
  emission lines is shown with redshift. The shaded regions enclose the
  ISAAC \j\ and \k\ bandpasses and the dot-dot-dot-dashed line shows
  the redshifted 4000\AA\ break seen in galaxies with evolved stellar
  populations. The solid lines refer to (clockwise) H I Ly$\alpha$
  \l1216, CIV \l1549, CIII]\l1909, Mg II \l2798, \oii\l3727, \hb\
  \l4861, \oiii\l5007, \ha\ \l6563, \nii\l6584, \sii\l6731,
  \siii\l9069, \siii\l9531, He I Pa$\delta$ \l10830, \pab\ \l12822 and
  \paa\ \l18756. See \S\ref{section:discussion} for discussion.}
  \label{fig:redshiftedlines}
\end{center}
\end{figure*}

\noindent
In this work we study a small sample of hard serendipitous sources in the near infrared (NIR) with the ISAAC instrument on the Very Large Telescope (VLT). There are a number of reasons for choosing the NIR regime: 1) as we have previously shown (Crawford et al. 2001), optically-dim sources are relatively bright in the NIR, making their {\sl study} possible; 2) XRB synthesis models have so far predicted a characteristic source redshift of $z=2$, where typically strong emission lines such as \ha\ or \hb\ would be redshifted to the NIR (Fig~\ref{fig:redshiftedlines}); 3) longer wavelength rest-frame lines 
are less sensitive to reddening and should be detectable through higher obscuring columns.

This extends our previous work of obtaining UKIRT spectra (Crawford et al. 2001) of such sources to deeper equivalent width limits and provides constraints on the utility of using an 8m class telescope in the NIR for such studies.

Quoted cosmological quantities assume
H$_0~=~50$~km~s$^{-1}$~Mpc$^{-1}$ and q$_0=0.5$ throughout.

\section{Sample Selection}

We are studying unresolved X-ray sources found serendipitously
in \c\ ACIS-S observations of nearby galaxy clusters,
preferentially selecting hard sources with count ratios $\ltsim$2 (see
section~\ref{sec:xray}). Roughly one-third of the entire ACIS sample
is hard by this definition.

Our total sample consists of more than 20 clusters and covers the
entire range in right ascension. Thus, our initial follow-up catalogue
is the Digitized Sky Survey
(DSS)\footnote{http://archive.stsci.edu/cgi-bin/dss\_form}, on
which we target the sources that are dim or invisible
($B$$\gtsim$22.5; $R$$\gtsim$21; hereafter referred to as
optically-dim). At the flux level that we sample, roughly 40 percent
 of all X-ray sources satisfy this criterion. As our pointings lie in
 the directions of massive galaxy clusters, we can make use of rich
 archival resources and often get fainter detections and/or limits.
We find that optically-dim sources in 10--20 ksec
ACIS exposures cover a wide range in magnitude: 22$\sim$$B$$\gtsim$24
and 19$\sim$$I$$\gtsim$23. Most of these are detected in the NIR at
16.5$\ltsim$$K$$\ltsim$19. For deeper \c\ exposures, our faintest
limits are $B$$>$25.5, $I$$>$24 and $K$$>$20.

Our \lq best\rq\ targets are the hard sources that are also optically-dim
and we estimate these to be about 50 percent of the
hard X-ray serendipitous sample. We also include a few exceptionally
hard sources which may have optical counterparts on the DSS.

For the present work, we selected 8 sources found serendipitously in 3
clusters observable from the southern hemisphere (see next section). 6
of these are hard, and 2 soft sources are included for comparison. 6
are optically-dim and 1 soft source has $R$=20. We were forced to
observe a relatively bright source with $R=19.6$ (but very hard) due
to poor atmospheric transparency on one night. This small sample
represents about 13 percent of all the detected X-ray sources, and 25 percent
of the hard sources.


\section{X-ray observations}
\label{sec:xray}
The \c\ cluster observations from which we chose sources for the present work are : MS\,2137.3-2353
(sequence number 800104, duration 50~ks), A\,1835 (800003 of 19.6~ks) and 
A\,2204 (800007, 10 ks). 
The effective exposure of the \c\ observation of
MS2137.3--2353 was limited by background flaring to 34.7~ks; the other
observations were unaffected by background variations. The data were
processed with the Chandra Interactive Analysis of Observations
(CIAO\footnote{http://cxc.harvard.edu/ciao/}) software, using versions
2.1 and 1.5. The Galactic line-of-sight column densities in the
directions of MS\,2137.3-2353, A\,1835 and A\,2204 are approximately
3.5, 2.3 and 5.6$\times10^{20}$cm$^{-2}$ respectively (Stark et al
1992).

The source detection was performed in the total 0.5-7~keV\ band in a
very similar manner to that described in Crawford et al (2002), using
the standard WAVDETECT package in CIAO. WAVDETECT was run on the data
in two forms: the original unbinned pixels (each of half-arcsec
width), and binned by two pixels (ie arcsec-long pixels). We used the
$\sqrt2$ sequence of wavelet scales (ie 1, 1.414, 2.0 \ldots 16.0
pixels), and set the significance threshold for sources at $10^{-6}$
(which implies that the expected number of false sources per ACIS chip
is roughly one\footnote{WAVDETECT manual at
http://cxc.harvard.edu/ciao/}). We also checked that the 8 sources
studied in this paper were detected with a lower threshold of
$10^{-7}$, thus making them unlikely to be spurious. We discarded all
sources within 20 arcsec of the edge of each chip to avoid loss of
source counts due to the spacecraft dither, and those with fewer than
ten (non-background subtracted) counts. On the ACIS chips from which
the present sample of sources was chosen, we found the following total
number of sources in the respective cluster fields. MS2137: 11 on
S2[chip 6] and 25 on S3[7]; A\,1835: 14 on S3[7] and 2 on I2[2];
A\,2204: 8 on S2[6] and 3 on I3[3].

We also estimated the counts in each source in 3 energy bands:
0.5-2~keV\ (soft; S), 2-7~keV\ (hard; H) and 0.5-7~keV\ (total). The
counts were taken from a box centred around a source, with a length
given by the square root of the number of pixels in the source cell
(as given from WAVDETECT). The local background was estimated from a
concentric box with a length five times longer. The background box was
offset if it would otherwise spill over the edge of the chip, or would
include a close neighbouring source. 

\begin{figure*}
  \begin{center}
\includegraphics[angle=90,width=8cm]{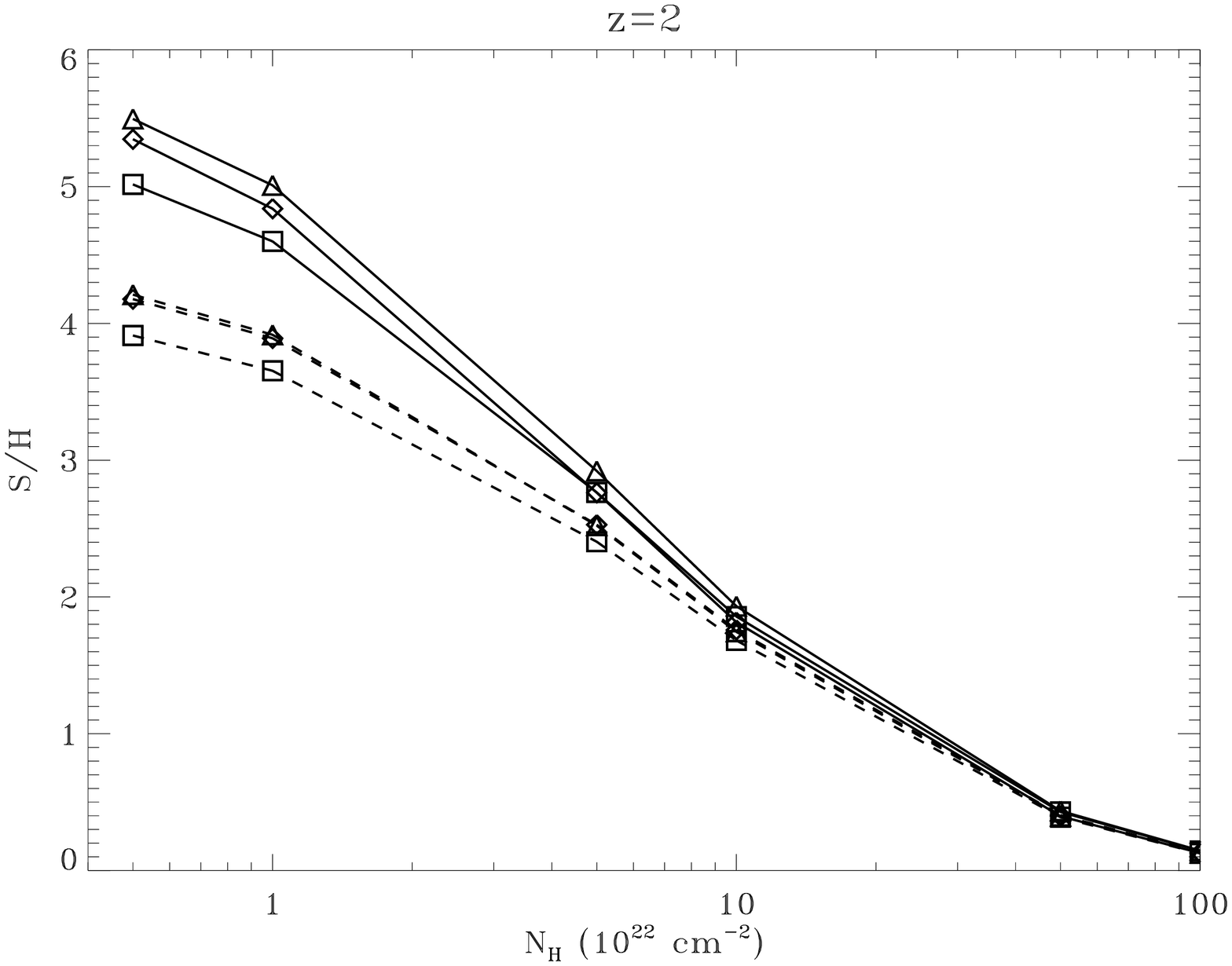}
\includegraphics[angle=90,width=8cm]{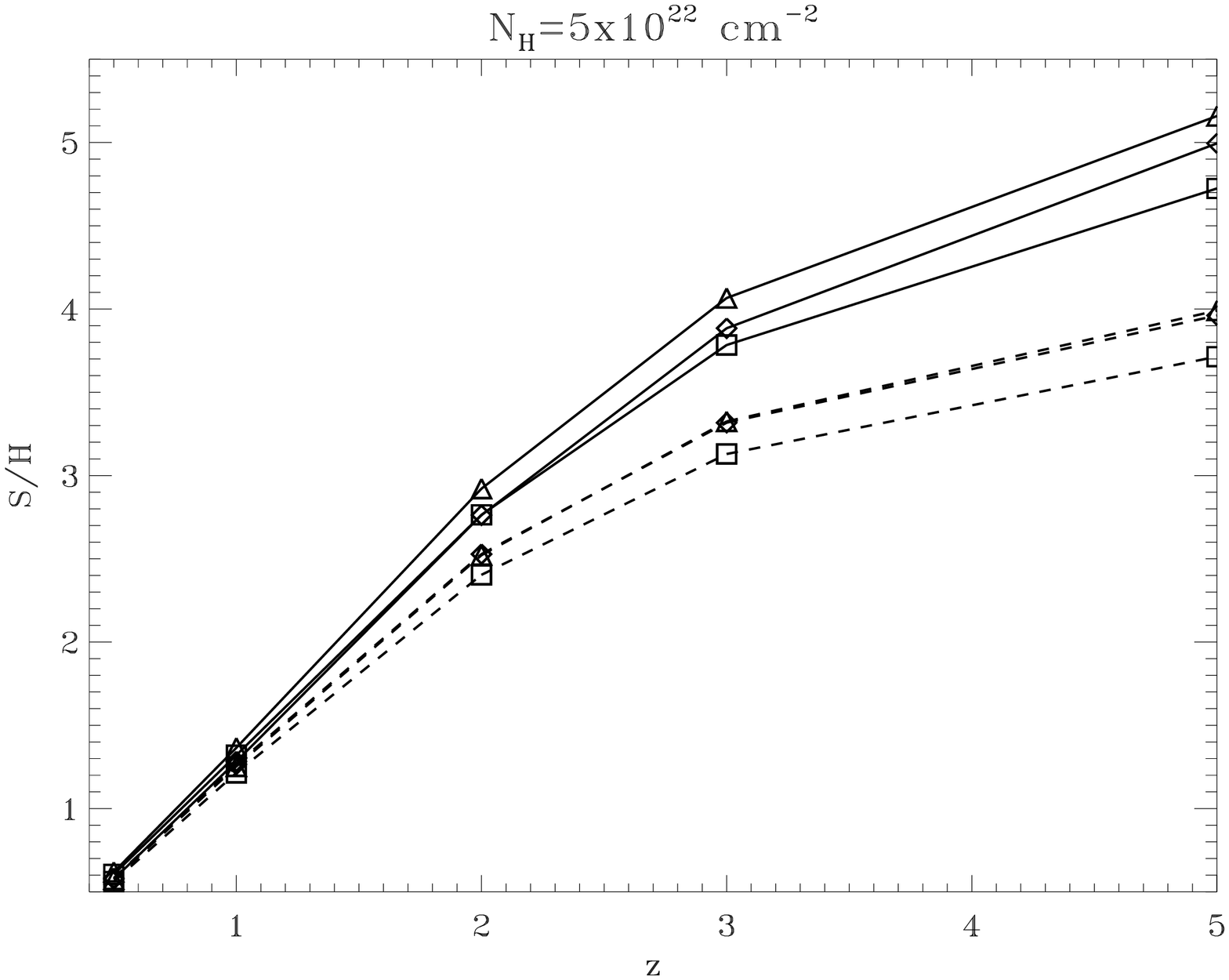}
  \caption{Predicted S/H ratios at various obscuring columns
  (Left, at $z=2$) and redshifts (Right, at N$_{\rm H}=5\times
  10^{22}$ cm$^{-2}$) for an absorbed power-law transmission model
  ($\Gamma =2$) and Galactic absorption appropriate to MS2137
  (diamonds), A\,1835 (triangles) and A\,2204 (squares) as measured on
  the ACIS-S3 background-illuminated chip (solid) and the ACIS-S2
  front-illuminated chip (dashed). } \label{fig:shratios}
  \end{center}
\end{figure*}

The X-ray properties of the
sources are listed in Table~1. None of these are magnified by strong lensing due the cluster potential. Six of the sources are very hard (with
S/H ratios of less than 1.6). Fig~\ref{fig:shratios} shows how the S/H ratio can be translated into physical column densities under different assumptions. A simple absorbed power-law ($\Gamma=2$) transmission model was adopted within XSPEC (Arnaud 1996) and counts in the soft and hard bands were predicted assuming different Galactic columns appropriate to the cluster line-of-sight and response matrices for each of the two kinds of
CCDs on the \c\ ACIS instrument. For the ratios that we observe (0.1--3; Table 1), the implied
columns are $5\times 10^{22} \sim 10^{24}$ cm$^{-2}$ if at $z=2$;
or the redshifts are consistent with $\sim 0.5<z<2$ if N$_{\rm
H}=5\times 10^{22}$ cm$^{-2}$. 

\begin{figure}
  \begin{center}
\includegraphics[angle=-90,width=8cm]{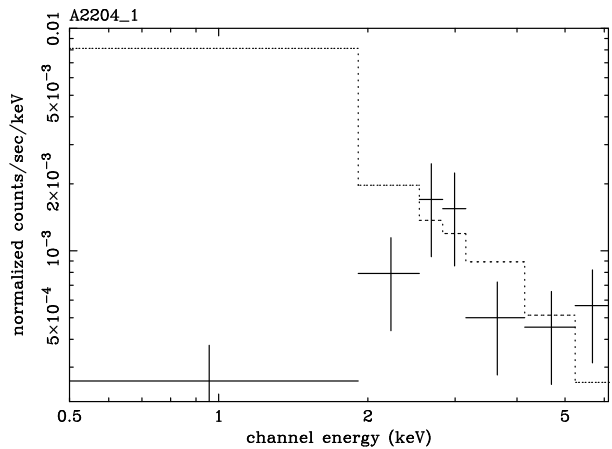}
  \caption{\c\ ACIS-S2 spectrum of A\,2204\_1. The 38 counts have been
  binned into groups with a minimum of 5 counts each. Though the
  errors are large, the plot clearly shows the hard nature of the
  source, with the least counts being detected in the soft band where
  a $\Gamma =2$ power-law model with only Galactic absorption predicts
  the most flux (dotted line).} \label{fig:a2204611xray} \end{center}
\end{figure}

Only two sources (A\,2204\_1 and A\,2204\_2) have more than 30 counts
in the total \c\ band. A2204\_2 is relatively soft with most counts
lying below 2~keV. A2204\_1 is hard and in Fig~\ref{fig:a2204611xray}, we 
show its extracted X-ray spectrum. Despite large errors, the
spectrum clearly demonstrates the need for large amounts of X-ray
absorption, as compared to the spectrum of a source emitting power-law
(photon-index $\Gamma =2$) radiation modified only by Galactic absorption
appropriate to the line-of-sight to A\,2204.

\section{Follow-up Imaging \& Photometry}

\subsection{Near-Infrared}
\begin{figure*}
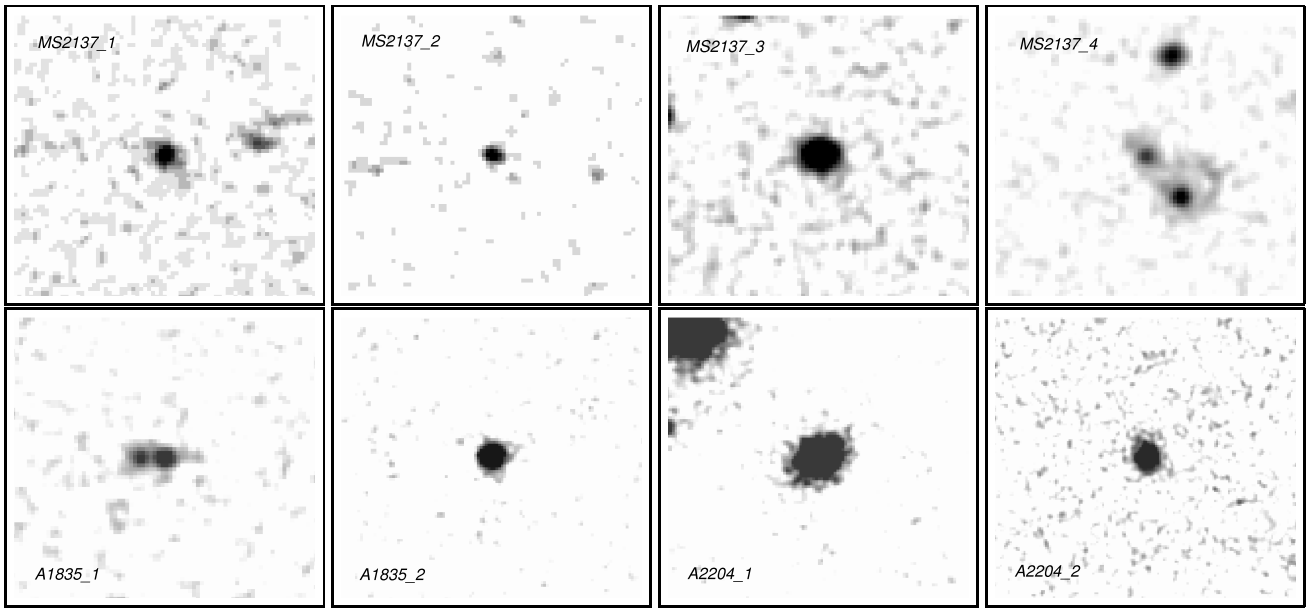

  \begin{center}
\fbox{\includegraphics[angle=0,width=4cm]{ms2137_1_K_named.ps2ps}}
\fbox{\includegraphics[angle=0,width=4cm]{ms2137_2_K_named.ps2ps}}
\fbox{\includegraphics[angle=0,width=4cm]{ms2137_3_K_named.ps2ps}}
\fbox{\includegraphics[angle=0,width=4cm]{ms2137_4_K_named.ps2ps}}
\fbox{\includegraphics[angle=0,width=4cm]{a1835_1_K_named.ps2ps}}
\fbox{\includegraphics[angle=0,width=4cm]{a1835_2_K_named.ps2ps}}
\fbox{\includegraphics[angle=0,width=4cm]{a2204_1_K_named.ps2ps}}
\fbox{\includegraphics[angle=0,width=4cm]{a2204_2_K_named.ps2ps}}
  \caption{\k-band smoothed thumbnails of all the sources. From left to right and top to bottom: MS2137\_1, \_2, \_3, \_4, A1835\_1 (all VLT ISAAC observations), A1835\_2, A2204\_1 and A2204\_2 (UKIRT UFTI). North is up and East is to the left, with each image being 10 arcsec on a side. In each case, the infrared source associated with the \c\ counterpart is the one closest to the centre of the image.
  } \label{fig:thumbnails}
  \end{center}
\end{figure*}

Near-IR images of the selected sources in the MS\,2137.3-2353 field
were obtained using the imaging spectrograph ISAAC (Moorwoord et
al. 1997) on the Very Large Telescope (VLT). Since the field of view
is 2.5 arcmin, only $\sim$1 \c\ source can be imaged at a time. Thus, imaging was targeted toward a small sample of hard sources (from which 4 were chosen on the subsequent nights for spectroscopy). We observed through the \j, \h\ and \k s (hereafter
\k) filters on 2001 June 28 in seeing of $\sim$0.3 arcsec (see
Fig~\ref{fig:thumbnails} for \k-band thumbnail images of all the
sources). The total integration time used was 600 s in \h\ and
\k, and 720--900 s in \j, with individual exposures of 10 s obtained
in jitter mode around a grid with offsets of $\sim$30 arcsec. Bad
pixel map creation, dark current subtraction and flat-field division
were carried out using the {\sl jitter} routine of the {\sl eclipse}
software package V4.0.4 (Devillard 1997).  In addition, {\sl jitter}
was used for background subtraction (using parameters suggested by
Iovino 2001) and combination of jittered frames.
Three photometric standard stars were observed
over the course of the night and the zeropoint in each filter was
found to be constant to within an RMS of 0.02 magnitudes.

Magnitudes were computed using the SExtractor package (Bertin \&
Arnouts 1996). We present the resulting magnitudes of our sources in
Table~2 and a log of exposure times in Table~\ref{tab:isaaclog}. These
are Kron magnitudes (with a Kron scale of 2.5; Kron 1980) for isolated
sources and seeing-corrected isophotal magnitudes (with minimum
isophote at 1.5$\sigma$) for blended objects.
It has been reported that the {\sl
jitter} pipeline may underestimate the brightness of sources in the
\k-band due to biasing of the local background in jittered images. We
corrected for this by a simple prescription of increasing all the
fluxes obtained at the VLT by 10 percent. Refer to Iovino (2001) for more
details.


The excellent seeing conditions enabled us to clearly resolve some
sources with a double morphology -- MS2137\_4 and A1835\_1 -- both possibly interacting (Figure~\ref{fig:thumbnails}).
The signal-to-noise decreases in the \h\ and \j-bands,
but the components are still resolved and separable for
photometry. MS2137\_3, A1835\_1, A2204\_1 and A2204\_2 show clear ellipticity in
the \k-band,
while the
remaining sources -- MS2137\_1, \_2 and A1835\_2 are unresolved.

The sources in the field of A\,2204 and A\,1835 were imaged at UKIRT
using the UFTI array (UT date 2000 Aug 11 and Feb 24 respectively;
A1835\_1 was also observed at ISAAC). A full description of the data
reduction is given in Crawford et al. (2001). Essentially, this was carried out using the standard package CGS4DR V2.001 in a manner very similar to that for the ISAAC data. Magnitudes were measured in SExtractor, and are presented in Table~2.

\subsection{Optical}
DSS (both generation 1 and 2) blue and red images of the X-ray fields were downloaded directly from the ESO mirror of the DSS website\footnote{http://archive.eso.org/dss/dss}. Deeper identifications of or detection limits to optical counterparts from the AAT, the CFHT and the INT were obtained from the respective archives. These data were calibrated using archival bias-subtraction and flat-fielding frames created on the night of observation. Fringing was removed from the $I$-band data by generating a master fringe frame from six offset frames for the AAT data. Table~\ref{tab:obsinfo} gives details of the archival datasets used.


Magnitudes were obtained using SExtractor
as described in the previous section. Photometric standard stars were used for flux
calibration (except one case; see Table~2). DSS magnitudes used a smooth extension of the first generation flux calibration by the Catalogs and Surveys Branch\footnote{http://www-gsss.stsci.edu/; based on observations by Lasker et al. (1988)} to fainter fluxes. Upper-limits for
non-detected objects were defined as the flux corresponding to 3 times
the background sky RMS in a 3-arcsec diameter aperture close to the source
location. Such an aperture size is typical of the Kron apertures for
the fainter of the detected sources.  The final magnitudes are listed
in Table~2. The INT Wide-Field Camera data was not fringe subtracted; thus apertures were selected by hand and background maps were inspected for consistency.

Typical seeing FWHM diameters for the optical observations were
$\approx$1~arcsec. Combined with a plate scale which is typically
larger than that of ISAAC, the morphology of the sources could not be
reliably determined in the optical.

\subsection{Source matching}
All fields were cross-calibrated with the APM sky survey catalogue\footnote{http://www.ast.cam.ac.uk/$^{\sim}$mike/apmcat/} to generate an astrometric solution for each. RMS errors for the solutions were less than a pixel (better than the pixel scale) in all cases over large image regions.

Any astrometric offset between a \c\ source and its corresponding NIR
counterpart is typically least ($\le 1$ arcsec) for sources on the ACIS-S3 chip 7,
closest to the telescope aim-point. With off-axis PSF
degradation, this may increase to several arcsec on the other
chips. For example, the centroid determination of A\,1835\_2 on ACIS
chip 2 has large errors (Table 1), which leads to an offset of 5
arcsec from the source that we consider to be the NIR
counterpart. There is, however, no source confusion problem
(Fig~\ref{fig:thumbnails}) and identification is unambiguous (see also \S~\ref{sec:a1835-2}). In all
cases, we could identify a NIR counterpart and associated it with the
nucleus of the object.

The probability of a false match occurring by chance was calculated by a \lq randomstep\rq\ method similar to that used by Hornschemeier et al. (2001). Astrometric cross-correlation with DSS images was repeated after all X-ray sources were offset 10 arcsec to the north-east, north-west, south-west and south-east. The number of false source matches, averaged over the four offsets, was small: $\le 0.5$ (zero-offset source match numbers ranged between 10 and 20). 

\section{Spectroscopic Observations}

We acquired spectra of the near-infrared counterparts to the \c\ X-ray
sources during the nights of 2001 June 29 and 30, again using ISAAC on
the VLT. We used a $1\times 120$ arcsec slit and the low-resolution
grating in the {\it SK} (hereafter \k) and/or \j\ bands. The choice of
the low-resolution grating was made since no apriori information on
the redshifts and detectability of the sources was known; the aim was
to capture a large amount of source flux through the slit. This
resulted in a dispersion of 7.138\AA\ per pixel and a full-width at
half-maximum (FWHM) of $\sim$45\AA\ for a single narrow night-sky OH
emission line assumed to be intrinsically unbroadened (R$\sim450$) in
\k, and a dispersion of 3.610\AA\ in \j, with a limiting
FWHM$\approx$25\AA (R$\sim500$). The first night had thick cloud at
times, with seeing varying from 1 arcsec to greater than 3 arcsec,
whereas the second night was photometric.

The objects were acquired by blind slit offsets from nearby bright
stars as measured in pre-imaging. In many cases, the slit was long
enough to encompass the star as well as target, thus providing a
constant monitor of target acquisition. The targets themselves were
nodded on the slit in an A--B--B--A pattern, with a small random
jitter offset about each of two nod positions. Typical total exposure
times of $\sim$2 hr were used, with each integration being 180
s long in order to be background limited.  Given the significant
loss of data quality on the first, non-photometric night, we only had time
to obtain spectra in the $J$ and $K$ wavelength regimes, covering
a wide choice of redshift space with two filters
only. 

Electrical ghost removal, dark current subtraction, distortion correction of sky lines, background subtraction, flat-fielding and combining of offset images was all performed using the {\sl eclipse} software, and more
specifically {\sl isaacp} packages written for the ISAAC pipeline. One-dimensional spectra were optimally-extracted with IRAF with a typical extraction aperture of 1.6 arcsec. A wavelength calibration solution was obtained by
cross-correlating observed night-sky emission lines
with a list of Rousselot et al. (2000).

Each target spectrum was divided by the spectrum of a 
spectro-photometric solar-analog star (reduced and extracted in an identical manner). Absorption features intrinsic to the star were removed by dividing
through a spectrum of the same spectral class (Pickles 1998) or a
Solar spectrum\footnote{The NSO/Kitt Peak FTS data used here for the
solar spectrum were produced by NSF/NOAO} which is at a resolution
similar to that of the star. Smoothing was performed with a
{\sl boxcar} of typically 5 pixels and finally, flux levels were
determined by scaling the count rate in the targets to the count rate
of the stars, whose magnitudes are known. The flux density as inferred
from the spectra matched well that inferred from imaging in most cases
(One notable exception was A2204\_1, observed through the slit during
cloud cover. The spectral flux density was lower by a factor of 2 than
the previous night's imaging measurement).

\begin{figure*}
  \begin{center}
   \includegraphics[angle=0,width=16cm]{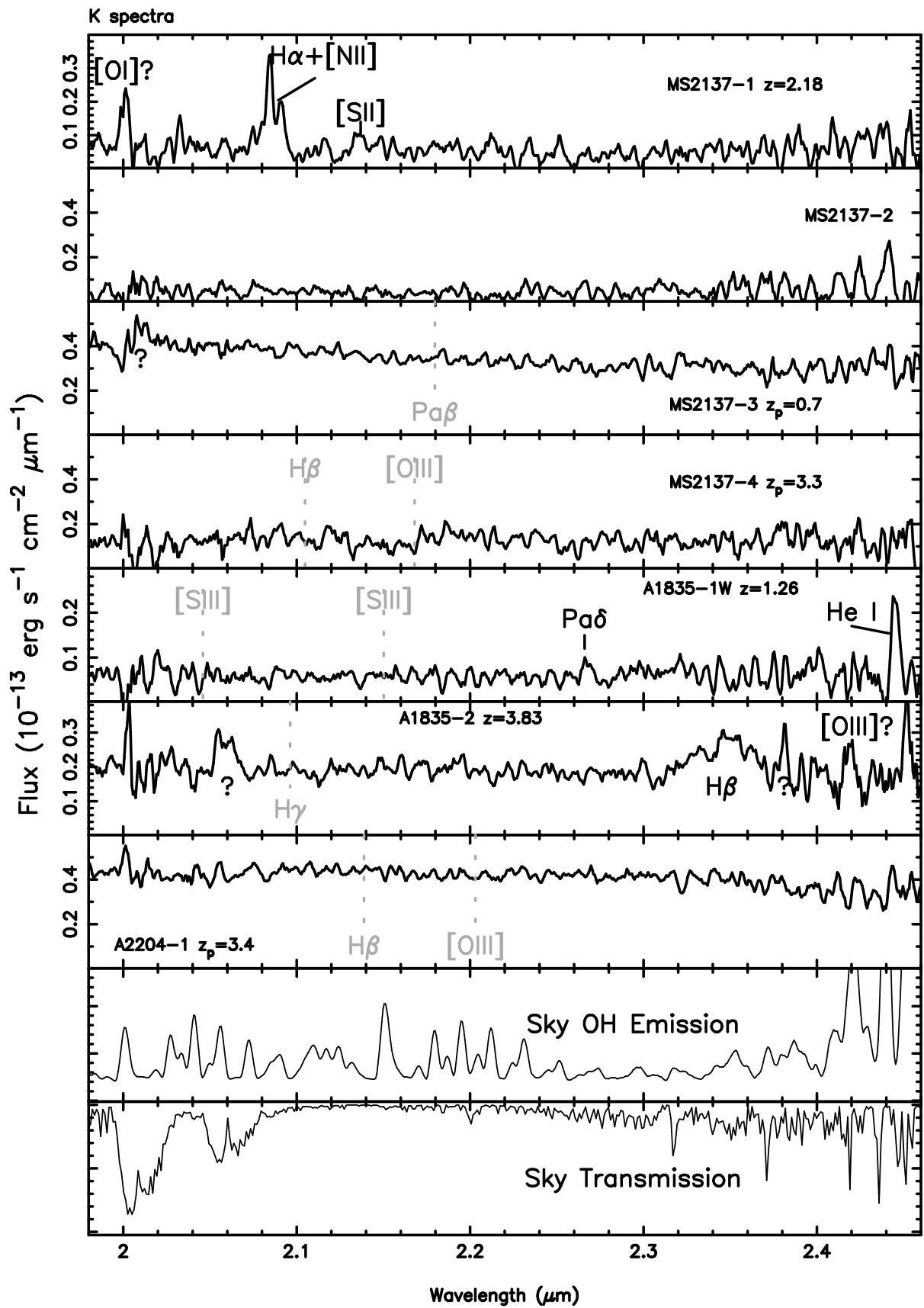}
  \caption{K-band ISAAC spectra. The y axis for all the target
  spectra begins at zero flux. The bottom two panels show the sky
  emission spectrum (at arbitrary scaling) and the sky transmission
  fraction (and hence the sky absorption). The sky emission spectrum
  has been effectively \lq flattened\rq\ by dividing a polynomial to
  fit the rising thermal continuum in order to enhance the lines
  themselves. Lines expected at the derived redshift (spectroscopic or photometric) are marked in light grey.
}  \label{fig:kspec}
  \end{center}
\end{figure*}

\begin{figure*}
  \begin{center}
\includegraphics[angle=0,width=16cm]{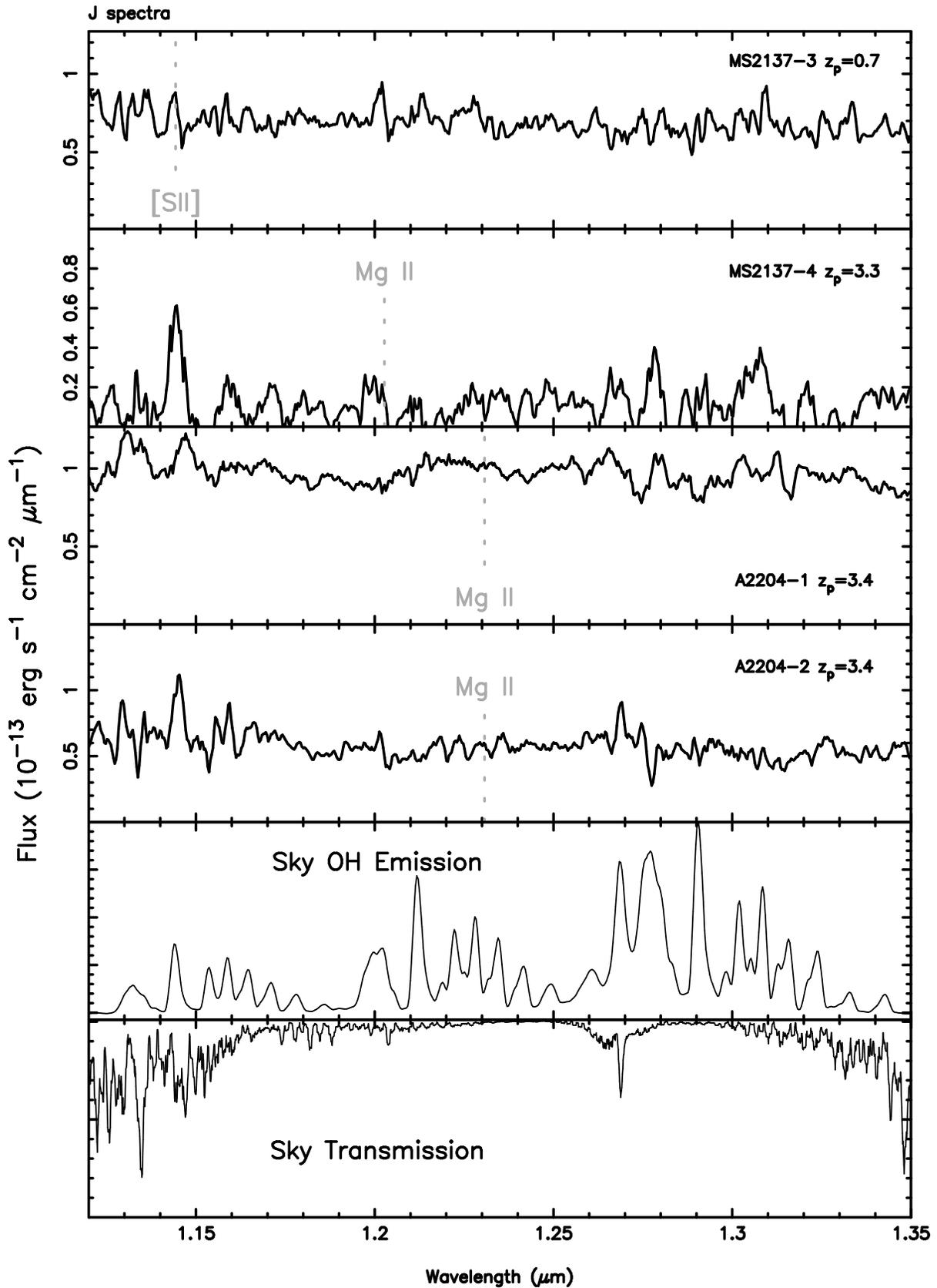} \caption{Same as
  in Fig~\ref{fig:kspec} for sources observed in the the J-band. The emission feature just before
  1.15\micron\ in all target spectra is most likely spurious (see text
  for details).} \label{fig:jspec}
\end{center}
\end{figure*}

The \k\ and \j-band spectra are presented in Figs~\ref{fig:kspec} and
\ref{fig:jspec} and a log of the spectroscopic observations with
exposure times and slit position angles is given in
Table~\ref{tab:isaaclog}.

\section{Results}

\subsection{Colours and photometric redshifts}
\begin{figure}
  \begin{center}
  \includegraphics[angle=90,width=8cm]{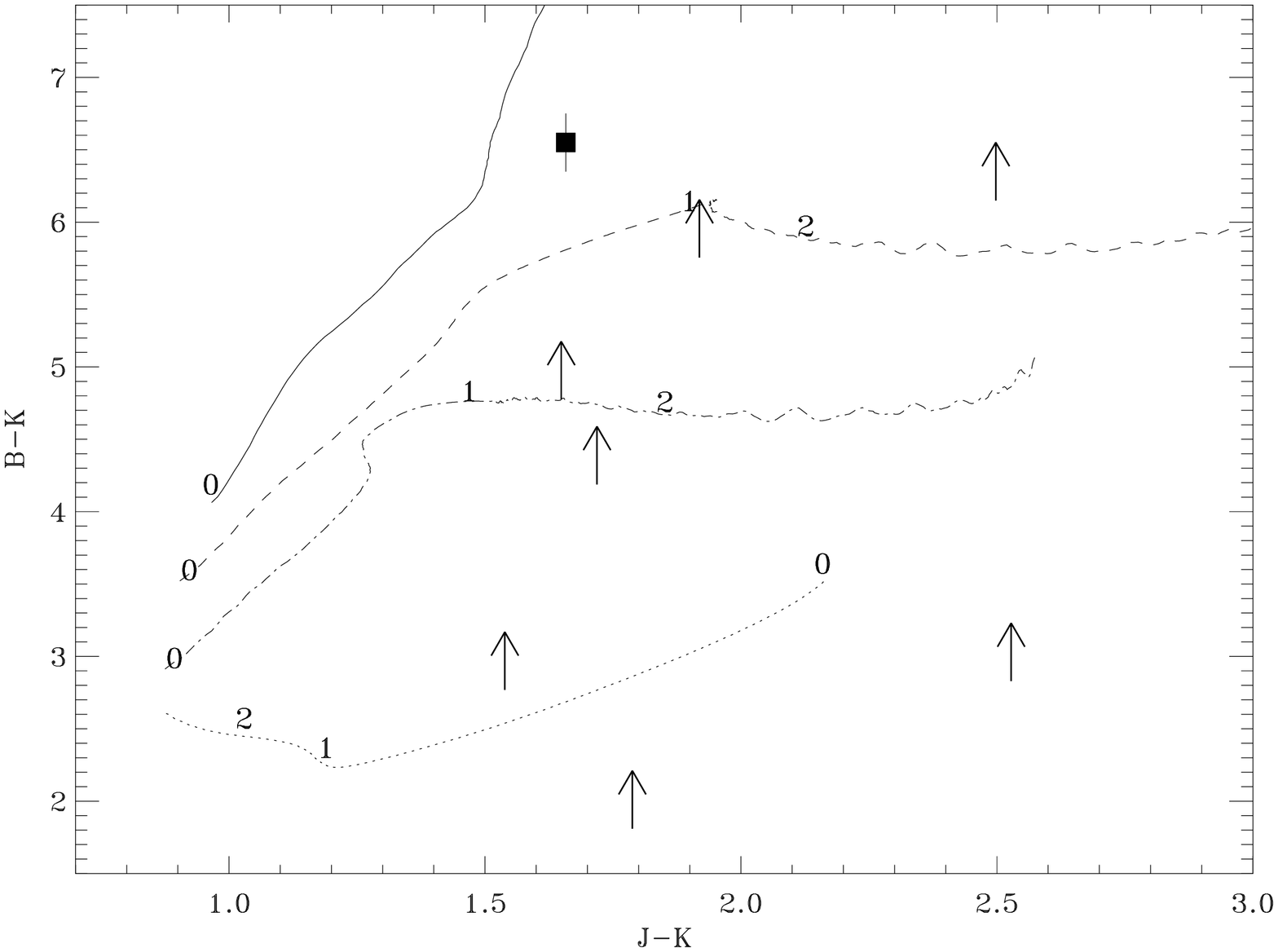}
  \caption{$B$--$K$ vs. $J$--$K$ colours and limits for the sources in
  this paper (filled square and arrows). All colours have been
  corrected for Galactic line-of-sight reddening. Colour tracks are
  shown for unreddened Coleman-Wu-Weedman (CWW) E (solid), Sbc
  (dashed) and Scd (dot-dashed) empirical templates. The dotted line
  is a colour track of a radio-quiet QSO from Elvis et al. (1994). The
  numbers associated with the lines mark the redshifts of the
  templates at those point.}
  \label{fig:bkjk} 
  \end{center}
\end{figure}

The hard S/H ratios argue against any of our targets being stars (The stellar locus is also well separated in $J-H$ vs. $H-K$ colours; c.f. a similar plot for sources in Crawford et al. 2002). 
The optical--infrared colours in Fig~\ref{fig:bkjk} show that most of
the sources have colours or limits which are redder than the $B-K$
colour of an unobscured quasar template (Elvis et al. 1994; median of
a sample of UV-bright, radio-quiet quasars at low redshift). The red
target colours are closer to those predicted for Coleman, Wu, Weedman
(1980) redshifted templates (shown without reddening).  The three
sources with the bluest $B-K$ limits (MS2137\_1, MS2137\_2 and
A\,1835\_1) are those with only relatively shallow optical
lower-limits from the DSS.

\begin{figure*}
  \begin{center}
\includegraphics[angle=90,width=8cm]{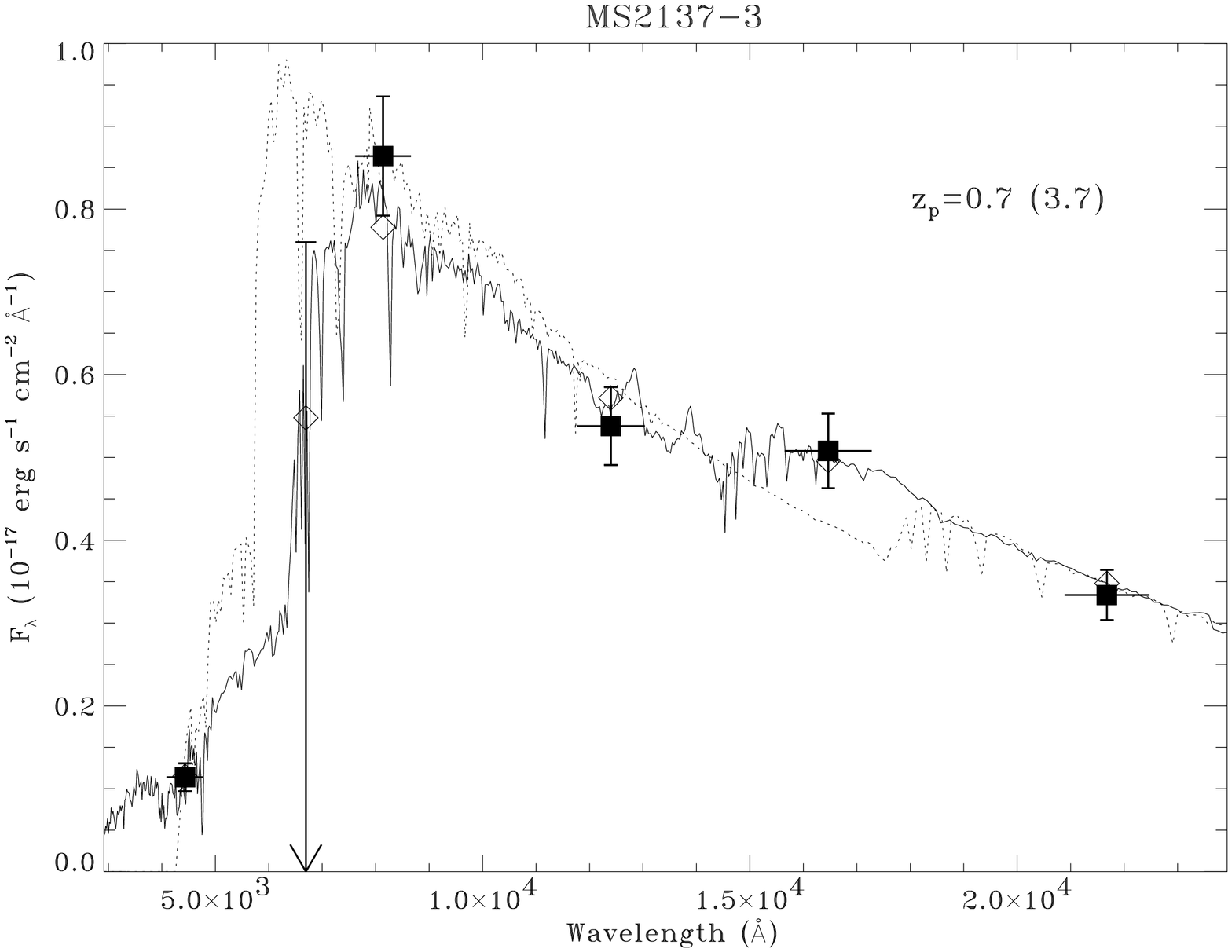}
\includegraphics[angle=90,width=8cm]{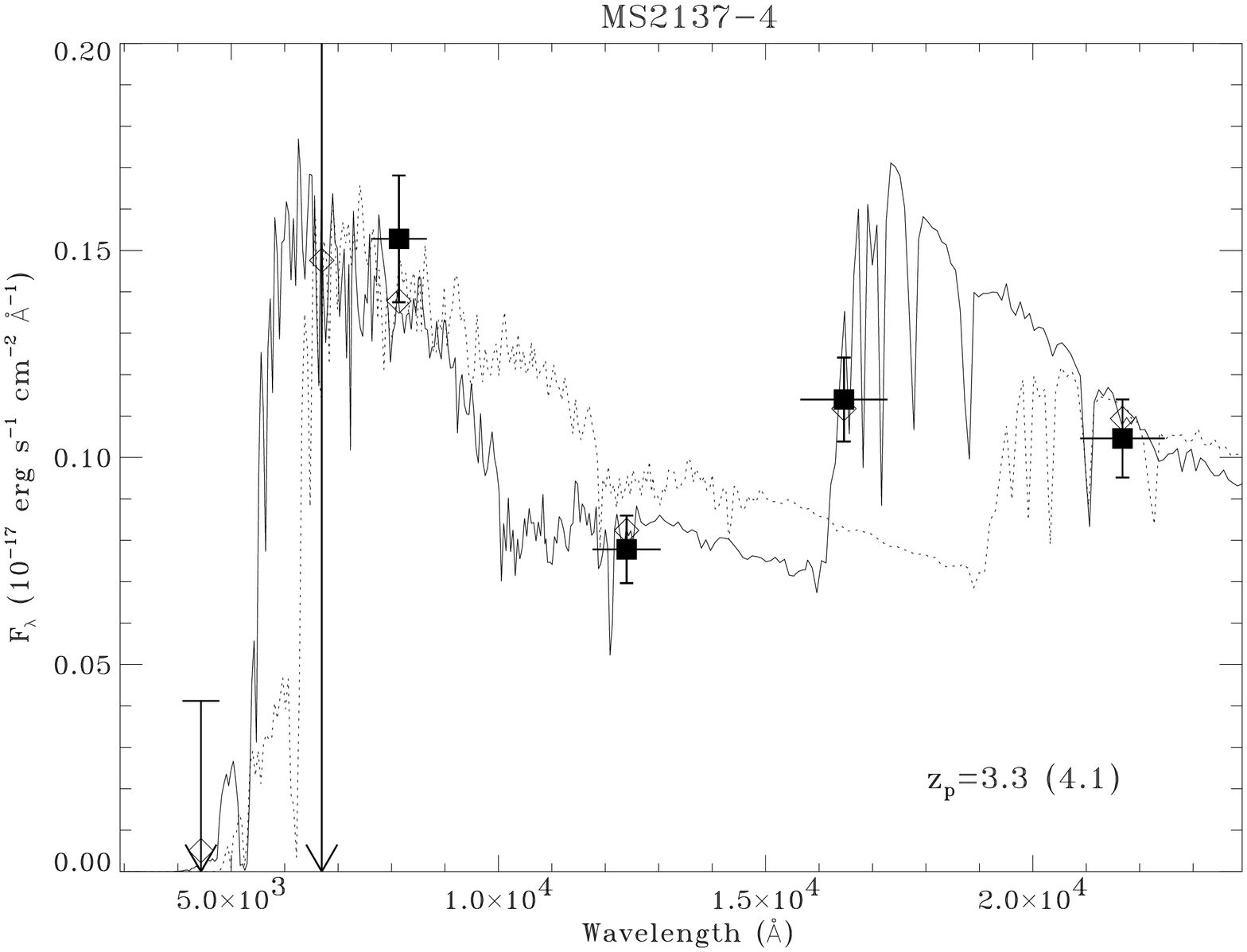}
\includegraphics[angle=90,width=8cm]{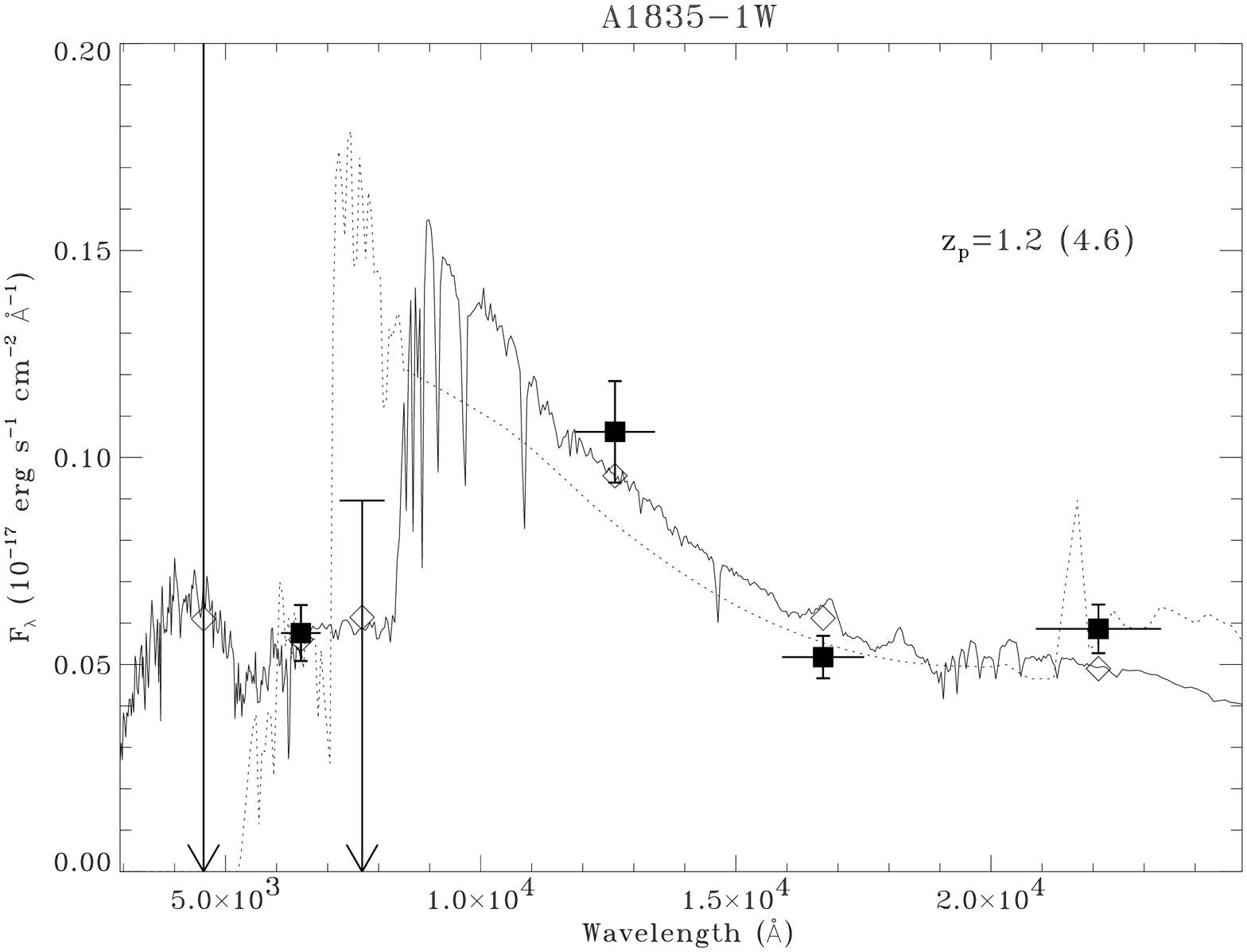}
\includegraphics[angle=90,width=8cm]{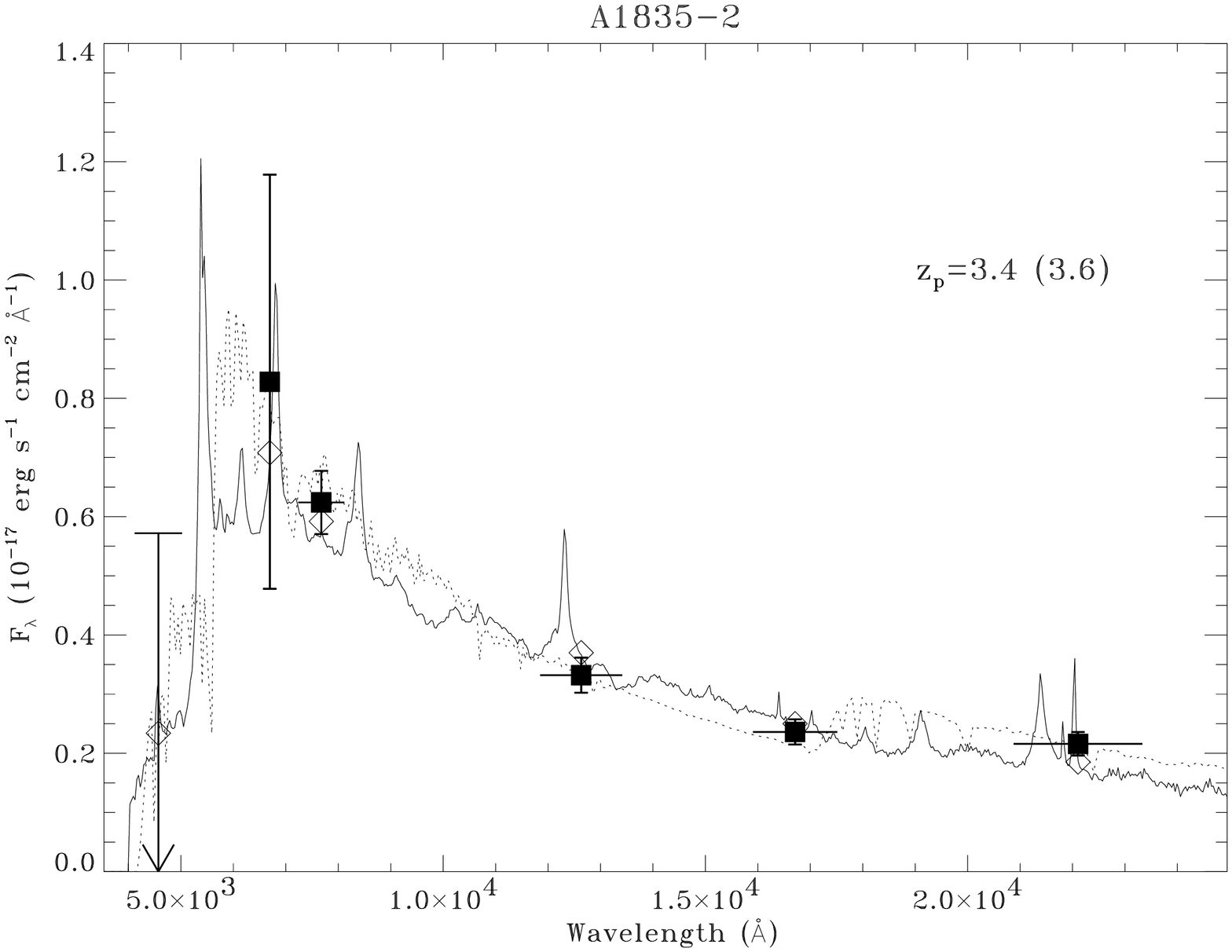}
\includegraphics[angle=90,width=8cm]{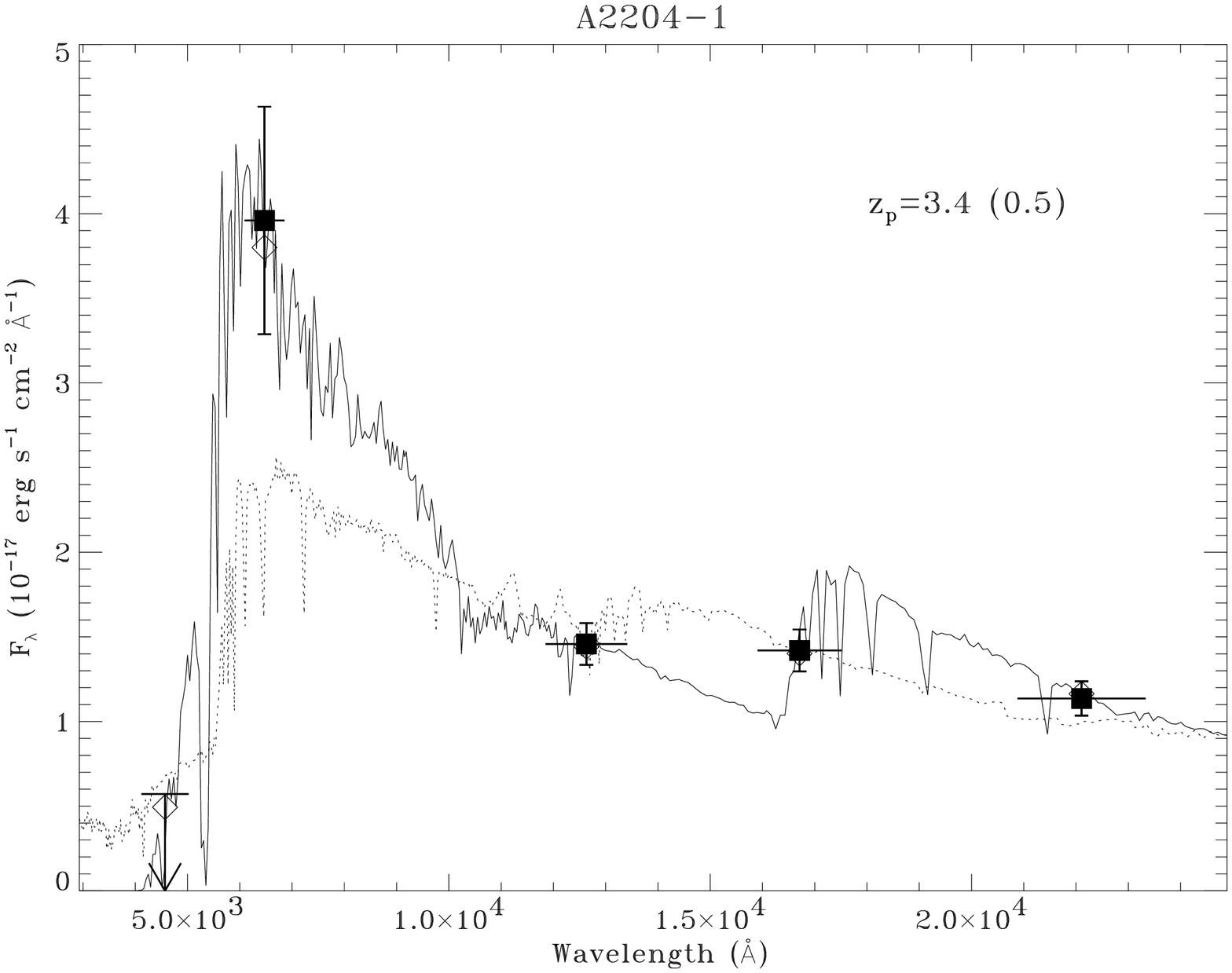}
\includegraphics[angle=90,width=8cm]{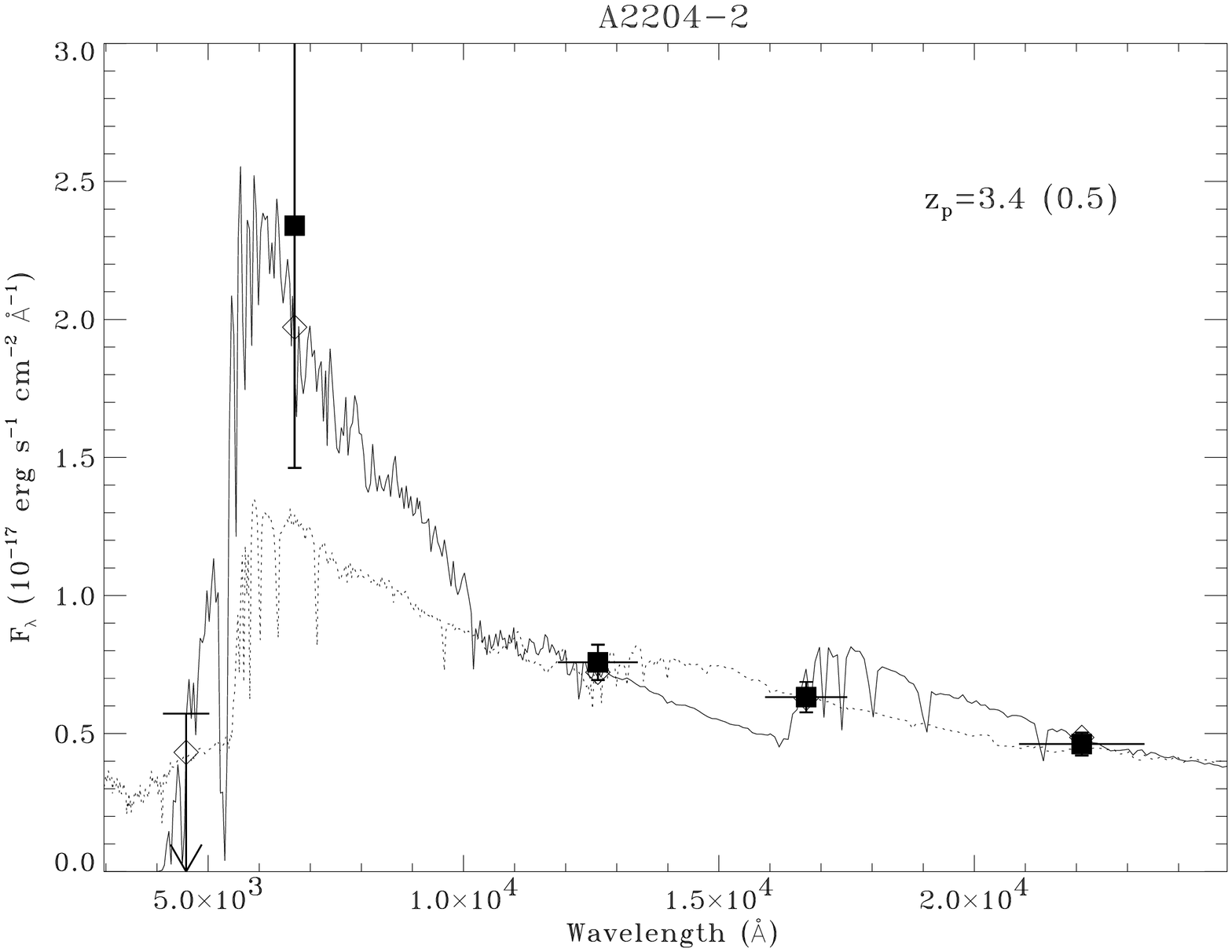}
\caption{The primary (solid) and secondary (dotted line) \zp\ solutions
fit to the observed fluxes and
limits (solid boxes and arrows) for 6 sources 
using HYPERZ. The unfilled diamonds mark the flux of the primary template
spectrum through the respective filters. The x-error bars denote the
approximate bandpasses and the y-errors are the 1$\sigma$ poisson flux
uncertainties. Labels denote the primary redshift solution, with the secondary solution in parentheses.  Refer to Table~\ref{tab:zphot} for more details.}
\label{fig:zphot} \end{center}
\end{figure*}

We estimated photometric redshifts (\zp) for all the sources using the
publicly-available code HYPERZ (Bolzonella, Miralles \& Pell\'{o}
2000), with the input parameters detailed in Crawford et
al. (2002). Briefly, HYPERZ varies the redshift and reddening of
various synthetic and empirical galaxy and AGN templates in order to
best fit the observed fluxes. The synthetic templates used are Bruzual
\& Charlot (1993) models which are also evolved with redshift. The
\zp\ solutions are listed in Table~\ref{tab:zphot}, and shown in Fig~\ref{fig:zphot}. 

If DSS upper-limits were shallow compared to the flux in the other filters, these were excluded from the HYPERZ fit (MS2137\_4 and A\,1835\_1). Note that 2 of the solutions (A2204\_1 and A2204\_2) are found at similar \zp$\approx 3.4$. This is primarily due to the large flux seen in the R--band relative to both B and J, placing the Lyman break at $z>3$. No reliable
redshift estimate could be obtained for the two sources (MS2137\_1 and
MS2137\_2) with DSS upper-limits much shallower than the deep \j, \h\
and \k\ magnitudes available.



\subsection{Spectra}

Three sources have strong detectable line emission features in their
near-infrared spectra; the rest have no ambiguous emission
(or absorption) features above the sky noise
(Figs~\ref{fig:kspec}, \ref{fig:jspec} and
Table~\ref{tab:lines}). Note that the apparent feature just red of
2\micron\ that shows up either as emission or absorption in many of
the spectra (Fig~\ref{fig:kspec}) is not real. Its wavelength overlaps
with that of the OH \l20005 and \l20008 8--6 Q emission transitions,
and also coincides with a strong telluric absorption feature. Since
the telluric spectrum is effectively divided out during standard star
division, a slight mismatch in the airmass between the target and the
star, combined with variation in the intensity of sky OH emission on
timescales of minutes, can produce such spurious features which are difficult to remove. Another such spurious
emission feature occurs at $\approx$1.145\micron\ in the \j-band
(Fig~\ref{fig:jspec}).

The limiting equivalent-widths (EW; observed frame) of the spectra
range from 20--60\AA\ in the \k-band and are close to 15--20\AA\ in
the \j-band, depending on the exposure time (except for MS2137\_4 in
the \j-band which was observed for 1800 s only and the limiting EW is
nearer 100\AA). This limit was estimated by measuring the equivalent
width of an unresolved gaussian profile with amplitude approximately 3
times the standard deviation of the continuum at various positions
along the dispersion axis. Note that this limit is shallower at the
positions of sky emission lines.

A discussion of the three sources with unambiguous emission lines follows, while notes for the remaining objects/spectra can be found in the appendix.

\subsection{Emission line sources}

\subsubsection{MS2137\_1}
\begin{figure*}
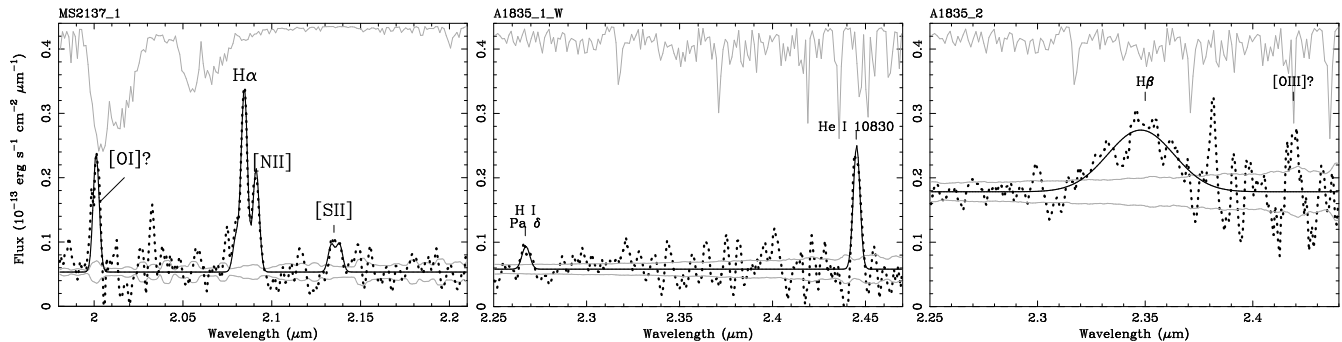

  \begin{center}
\includegraphics[angle=270,width=6.1cm]{ms2137_1_sky_abs_mod.ps}
\includegraphics[angle=270,width=5.7cm]{a1835_1_w_sky_abs_mod.ps}
\includegraphics[angle=270,width=5.7cm]{pgplot.ps}
  \caption{Zoom-in of the ISAAC \k-band spectra of the sources with
  strong emission lines -- MS2137\_1, A1835\_1 (the western component)
  and A1835\_2. The data is shown as bold dots, the fitted model with
  a solid line and the 1$\sigma$ errors due to the sky above and below
  a constant fitted continuum as the faint lines. Sky absorption has
  been rescaled (same for all figures) and is plotted as the faint
  lines at the top. The spectra have been smoothed over 5 adjacent
  pixels. \label{fig:zoom}}
\end{center}
\end{figure*}

MS2137\_1 is invisible on the DSS images, but identified as a
point source in the NIR.  Its \k-band spectrum (top spectrum
in Fig~\ref{fig:kspec}; Fig~\ref{fig:zoom} for detail) shows an
unambiguous \ha+\nii\ complex at 2.0846\micron, implying a redshift of
$z=2.176$\p0.001 for the source. Line emission from \sii\l\l6717,6731
at the same redshift is marginally detected (at 2$\sigma$ above the
sky noise). While it appears that \oi$\lambda$6300 is clearly detected,
caution must be applied as this coincides with the spurious feature just beyond 2\micron. 

A model with discrete symmetric Gaussian
profiles and identical velocity widths fixed to that of \ha\ overlaid
on a constant continuum was fit using QDP (Tennant 1991). The \ha\ line is unresolved and substantially narrower than broad type 1 AGN permitted lines.
The \nii\l6584:\ha\
line intensity ratio is 0.56 and \sii\l6717:\ha\ intensity is 0.12, which is 
consistent with that observed both in H II regions and
Seyfert 2s (Veilleux \& Osterbrock 1987; also Storchi-Bergmann 1991).


While observations of additional emission lines (e.g., redshifted
\oiii\l5007 in the \h-band) may help to distinguish between starburst
or AGN origin for the observed \k-band emission lines, the hard X-rays must
originate from a Seyfert 2. The implied X-ray (2--10 keV restframe)
luminosity, assuming a $\Gamma=1.4$ power-law at $z=2.176$ affected
only by Galactic absorption, is L$_{2-10}=5.6\times 10^{43}$
\ergps. If instead, we assume a crude model based on the observed S/H
ratio of $\approx$1.1, implying an obscuring column of $\approx
2\times 10^{23}$ cm$^{-2}$ for $\Gamma =2$ at $z\approx 2$
(Fig~\ref{fig:shratios}), the implied L$_{\rm 2-10}=1.6\times 10^{44}$
\ergps, after correcting for the obscuring column density.

\subsubsection {A\,1835\_1}
\label{section:a18351}
A\,1835\_1 is clearly resolved into two components on the ISAAC
\k-image (Fig~\ref{fig:thumbnails}). An astrometric solution
associates the brighter westerly component (A\,1835\_1W) with the X-ray source
(though only marginally). 
The ISAAC spectrum of this brighter 
component has a single strong emission line at 2.4452\micron\ , again unresolved (fifth spectrum from top in Fig~\ref{fig:kspec}). Although the night sky thermal
emission begins to dominate beyond $\sim 2.35$\micron, the narrow line is very significant (Fig~\ref{fig:zoom}) and lies fortuitously
between two bright OH emission lines. The line connects both components of the source and is thus spatially extended (we were able to obtain the spectra of both components simultaneously with a slit orientation of 90$^{\circ}$).


We attempted to identify weak features in the spectrum without much success, before trying to use the photometric redshift estimated from
HYPERZ as a first guide towards possible line identification.
\zp=1.2 [1.0, 1.5] (Table~\ref{tab:zphot}) suggests that the strong line at 2.4452-\micron\ 
and a weak $\sim$3$\sigma$ emission feature at 2.2673\micron\ could be He~I \l10830 and Pa$\delta$ \l10049
respectively (Fig~\ref{fig:zoom}), implying a tentative redshift
$z=1.256$, well within the 90 percent \zp\ confidence interval. 
The He I transition has been
observed previously in Seyfert 2s with intensity approaching \hb\ (and even
exceeding it in NGC 1068 and Mrk 3; Rudy et al. 1989; Osterbrock,
Shaw \& Veilleux 1990). Its strength is enhanced due to collisional
excitation from the metastable $2^3S$ level to $2^3P$, followed by
radiative decay (Osterbrock 1989). The other strong feature expected in
this regime (\siii\l9531) is not observed by us. This
may imply that the density of the narrow-line region is much greater
than the critical density to collisional de-excitation
N$_{critical}\approx 6\times 10^{5}$~cm$^{-3}$. If our identification
is confirmed, this would be the highest redshift observation of He I
\l10830 to date.


Estimating the intrinsic luminosity 
by correcting for a 
column-density of $\approx 10^{23}$~cm$^{-2}$ as implied by the S/H
ratio of 0.41 (on a front-illuminated chip) at the inferred redshift, we deduce the presence of a
powerful Seyfert emitting power-law radiation ($\Gamma=2$ fixed) with
L$_{2-10}=1.2\times 10^{44}$~\ergps.

\subsubsection{A\,1835\_2}
\label{sec:a1835-2}
The large $\sim$5 arcsec offset between the \c\ source A\,1835\_2 and its counterpart is most likely due to the difficulty of centroiding with off-axis ($\approx$12 arcmin) PSF degradation. The formal probability of a match at such a large offset is 1\%. However, if this were not the counterpart, a large X-ray:optical flux ratio and corresponding optical/NIR extinction would be inferred since the counterpart would be invisible in our images. This is unlikely given the softness of the X-ray source.

The spectrum (sixth from top in Fig~\ref{fig:kspec}) -- is not easy to
interpret. There is a clear broad (FWHM$\sim$4600 km s$^{-1}$) feature
centred at 2.3480\micron\ and two ambiguous features marked with a \lq?\rq\ in the figure.
No obvious emission line pattern fits
these data, and it is likely that the ambiguous lines are spurious and caused by telluric absorption in the first case, and increased thermal noise in the second.

The presence of a broad feature (Fig~\ref{fig:zoom} for detail), the high S/H ratio (Table~1) and the point-like morphology
(Fig~\ref{fig:thumbnails}) suggests that we are viewing a
quasar. Fitting a QSO template to the broad-band colours gives an
acceptable fit at \zp=3.4 [3.0, 4.0] (Table~\ref{tab:zphot}). Thus a
potential match for the broad feature
is with \hb\ for $z$=3.830\p0.005 (within the 90 percent \zp\ interval). Any
\oiii\l5007 emission would overlap with an OH emission feature close
to 2.1488\micron\ (refer to Fig~\ref{fig:kspec}; if this \oiii\ line
exists, it must have EW$<$40\AA\ in the observed frame). 
We do not see any \hg\l4340 emission. Its predicted photoionization
recombination intensity is 0.47 times that of \hb, though
we note that Vanden Berk et al. (2001) found that EW(\hg) $\approx
0.25\times$EW(\hb), which implies an \hg\ strength close to our
limiting EW (Table~\ref{tab:lines}). Moreover, any reddening is likely to further decrease the \hg:\hb\
intensity ratio. Thus, its absence may not be surprising.

Again, assuming $\Gamma=2$ and correcting for a column-density of
$\approx 7\times 10^{22}$~cm$^{-2}$ (S/H = 3.1), the implied
L$_{2-10}\approx 6\times 10^{44}$~\ergps at $z_{\rm tentative}$=3.830,
which is consistent with quasar luminosity.
%
\section{Discussion}
\label{section:discussion}
Of the 8 objects, we detect a broad \k-band emission line in one source (A\,1835\_2) and
narrow lines in two sources (MS2137\_1 and A\,1835\_1). The spectrum of the narrow-line source MS2137\_1 places a firm redshift constraint of $z=2.18$. The other two sources have only a single unambiguous line in their spectra and tentative identifications (at $z=1.26$ and 3.8) for these lines are determined in conjunction with the photometric redshifts. Two sources (MS2137\_3 and A2204\_1) have high signal:noise \j\ and \k-band featureless continua. The limiting observed-frame equivalent-widths (EW) for these are $\sim$20\AA\ in both bands. For the remaining sources, the limiting EW range over $\sim$30--60\AA\ in the \k-band. Few useful constraints can be obtained from the noisy spectra of MS2137\_2 and MS2137\_4 (\j-band). 

The high quality NIR images reveal that the sources have a mixture of unresolved (MS2137\_1, MS2137\_2 and A1835\_2), galaxy-like (MS2137\_3, A2204\_1 and A2204\_2) and double (MS2137\_4 and A1835\_1) morphologies.

To investigate the lack of significant spectral features in some sources, we ask the question: how likely is it that an emission line from a Seyfert galaxy at a randomly chosen redshift would be redshifted into the $J$ or $K$ bandpasses? Fig~\ref{fig:redshiftedlines} goes some way in answering this question. Most of the transitions mentioned in the figure are strong lines in Seyfert spectra, and all can easily have an EW greater than our best limiting value of 20\AA. There are few regions in redshift space where no redshifted lines are observed in either (or both) of the ISAAC bandpasses.
Especially important is the redshift regime $z\ltsim 1$, as
suggested by recent \c\ and \xmm\ findings (e.g., Rosati et al. 2001,
Brandt et al. 2002, Hasinger 2002). 
If the sources are at $z\ltsim 1$, we would see lines ranging from \ha\l6563 to \paa\l18756.
Table~\ref{tab:z1lines} lists these lines and their typical strengths. \ha\ and \nii\ are the strongest expected lines, followed by He I and forbidden transitions of Sulphur. 

In summary, at $0.7<z<1$, we would observe \ha$+$\nii. Detection of other (typically weaker and/or lower $z$) lines is also likely, as they have been observed in the literature with large strengths and EW.
Thus, although we cannot rule out the possibility that we miss the lines at $z<1$, we consider alternative possibilities as well.

Early-type galaxies with little on-going star-formation show many absorption features in their spectra (that we would not detect due to signal:noise constraints) but few (if any) emission features. This is especially true for extremely-red objects (EROs). In a sample of ISAAC spectra of galaxies selected for their extremely red colours, Cimatti et al. (1999) found neither strong emission lines nor continuum breaks. While it may be the case that our NIR spectra are dominated by galactic light, the hard X-ray luminosities (assuming redshifts as determined from photometric or spectroscopic constraints) range over $10^{43}-6\times 10^{44}$ \ergps. The X-rays must, therefore, originate in an active Seyfert-like nucleus, and the absence of any optical/NIR lines associated with the AGN must be accounted for. We consider the possibility that large column density gas and associated dust scatters / absorbs these line photons.


As discussed in section~\ref{sec:xray}, the hard S/H ratios (Table 1) suggest high obscuration of the nucleus.  
Intrinsic columns below $\sim 10^{22}$ cm$^{-2}$
will result in S/H ratios which are softer than observed ($\gtsim 3.5$) for all $z\ge 1$, if $\Gamma=2$. 
The 6 hard sources must have N$_{\rm H}>5\times 10^{22}$ cm$^{-2}$ if at $z=2$ or
N$_{\rm H}>10^{22}$ cm$^{-2}$ ($A_V>6$ mags assuming a Galactic dust:gas ratio) if at $z=0.5$. We note that evidence for such a dusty
environment was found for at least one type 2 QSO in the field of
A\,2390 by Wilman, Fabian \& Gandhi (2000) through radiative transfer
modelling based on photometric detections in the ISOPHOT mid-IR bands
(cf. Crawford et al. 2002). The column density of the gas as inferred
by the optical depth of the dust was found to be consistent with the
X-ray measurements. A model incorporating dust in narrow-line region clouds was proposed by Netzer \& Laor (1993).

The very hard ratio of 0.16 for A2204\_1 (a source with high
signal:noise featureless continuum) suggests a column $>10^{23}$
cm$^{-2}$ for all $z\ge 0.5$. In fact, extrapolating the de-absorbed
flux at 1 keV for a $\Gamma=2$ power-law model to $B$-band flux
assuming a \lq typical\rq\ quasar broad-band energy distribution
(i.e. $\alpha_{BX}$ spectral index; Elvis et al. 1994) implies $B_{\rm
predicted}\approx 19.5$ if A2204\_1 lies at $z=0$, or $B\approx
20.2$, if at $z=2$. For the observed DSS limit of 22.5,
this implies at least 2.5 optical magnitudes of extinction to the nucleus.

Compact, very optically-thick obscuration of the kind proposed by Pier \& Krolik (1992) will obscure emission from the nucleus itself. If obscuration is in the form of moderately thick tori spread over tens or hundreds of parsecs (Granato, Danese \& Franceschini 1994), line emission from star-forming regions could be partially absorbed as well. However, the distinct lack of emission lines (in both the optical and NIR), in these and other recently-found hard X-ray sources, requires covering fractions that are close to 4$\pi$. As opposed to low-redshift seyferts, obscuration at high redshift would then be mostly independent of orientation. Indeed, large covering fractions (85 percent) have been predicted by Fabian \& Iwasawa (1999), by correcting for absorption in the spectrum of the XRB.
  Forthcoming observations in the far-IR will
significantly improve our understanding of the scale and extent of the
enshrouding gas/dust distribution (e.g., SIRTF; Brandl et al. 2000).



\section{Conclusions}

We have obtained VLT ISAAC spectra of 8 \c\ X-ray
sources and detect continuum emission from all. These include 7
spectra in the \k-band and 4 in \j. 6 of these are optically-dim and the remaining 2 are dim in the $B$-band. The X-ray spectral count ratios
constrain the obscuring column; 6 sources are very hard with the
hardest being consistent with an intrinsic obscuring column density
N$_{\rm H}\gtsim 10^{23}$ cm$^{-2}$. These are typical of the
population which contributes the maximum flux per source to the 
X-ray background.

We have been able to identify 2 narrow-line AGN (one at $z=2.18$ and
one possibly at $z=1.26$; both consistent with luminous Seyfert 2s)
and one broad-line AGN possibly at $z=3.83$ for which intrinsic quasar
luminosity is inferred. This includes possible detection of the most
distant He I \l10830 emission to date. Spectra in other wavebands will
help to confirm these. Photometric redshifts have also been determined.

Even in the above long integrations ($\sim$ 1 or 2 hours long) on an
8-m telescope, we are able to detect significant emission lines in
only 3 of the 8 sources and identify the redshift of one of these
unambiguously. This extends our previous 4-m UKIRT observations to
better equivalent-width limits and confirms our earlier findings that
although such sources are readily observed in the near-infrared,
detailed identification of the source type is not straightforward. The
absence of emission lines in many type 2 AGN, which contribute a large
fraction of the XRB intensity, is evidence for high covering
fractions of intrinsic obscuring gas.

\section{Acknowledgements}
The work presented here is based on observations obtained with the \c\
telescope and the VLT. In addition, use is made of archival imaging
data from a number of telescopes (the AAT, INT, WHT and CFHT) and the
DSS. It is a pleasure to acknowledge the support received from all
these organizations in the process of acquiring our data. The {\it
eclipse} team at ESO is thanked for constant help with ISAAC data
reduction. We are grateful to the referee for thorough and constructive
criticism. PG would like to thank Andrew Bunker for useful discussions
and the Isaac Newton Trust and the Overseas Research Trust for
support. CSC and ACF acknowledge financial support from the Royal
Society.

\section*{Appendix: Notes on sources without significant emission lines}

\subsubsection*{MS2137\_2}
By an appropriate orientation of the long slit, MS2137\_2 (second
spectrum from top in Fig~\ref{fig:kspec}) was observed simultaneously
with MS2137\_1. Unfortunately, the only optical photometric data
available to us is the DSS, on which, like MS2137\_1, the source is
invisible. Compared to the deeper ISAAC $J$, $H$ and $K$ detections of
this source, the $B$ and the $R$ upper-limits are too shallow to
obtain a satisfactory photometric redshift solution. The source is
very dim and point-like in $K$, with no obvious spectral features
down to a 3$\sigma$ limit on the EW of $\sim 40$\AA.

\subsubsection*{MS2137\_3}
MS2137\_3 (third \k\ spectrum from top in Fig~\ref{fig:kspec} and top spectrum in Fig~\ref{fig:jspec}) is the
only source to show an obvious slope in the continuum, which rises
gently toward the blue in the K-band and drops suddenly at
$\sim$2\micron. Though reminiscent of a P-Cygni profile, the proximity
of the \lq break\rq\ to the OH emission at 2.0008\micron\ as well as
the deep telluric absorption strongly suggests that this drop is due
to the sky. 
The signal:noise in the continuum is $\sim 10$ per pixel, which may
also explain the lack of significant absorption features. (The typical
depth of such features expected at $z\approx$ \zp\ in the \k-band
is also about 10 percent of the continuum; e.g., the temple NIR galaxy spectrum of Mannucci et
al. 2001)


\subsubsection*{MS2137\_4}
MS2137\_4 lies in an apparently interacting system
(Fig~\ref{fig:thumbnails}) of galaxies with very red $B-K$ colour and diffuse tidal structure in
the $K$-band image. Astrometric calibration
associates the X-ray source with the fainter, more north-easterly
component.  The signal:noise in the continuum spectrum (fourth from top in Fig~\ref{fig:kspec}) is
comparable to that of MS2137\_1; however, unlike MS2137\_1, we do not
detect any emission features, though \hb\ would lie in the \k-band, if
$z=$\zp=3.3. The source is then red in $B-K$ due to the $k$-correction
of the Lyman break (Fig~\ref{fig:zphot}). Due to time constraints, only a 30-minute exposure
could be obtained in the \j-band (fourth spectrum from top in
Fig~\ref{fig:jspec}). This is also the faintest \j-band source
observed spectroscopically, which implies a poor equivalent width
lower-limit EW$^{3\sigma}_{\rm lim}\approx 100$\AA\ for any real
features.

\subsubsection*{A\,2204\_1 and A\,2204\_2}
A\,2204\_1 has a flat, featureless spectrum in both the \j\
and \k\ bands (seventh spectrum from top in Fig~\ref{fig:kspec} and
third spectrum from top in Fig~\ref{fig:jspec}), despite being the
brightest NIR source in our sample (It is also the second brightest
and the hardest X-ray source; Fig~\ref{fig:a2204611xray} for the X-ray
spectrum). Unfortunately, the ISAAC spectra of this source were
observed on a non-photometric part of the night with seeing $\sim3$
arcsec, which led to severe degradation of the signal:noise, thus
precluding possible identification of absorption features (The
1$\sigma$ photon noise variation is $\approx 8$ percent of the continuum in
the best part of the spectrum close to 2.2\micron). Intriguingly, the
blue $R-J=1.3$ colour predicts a relatively young system
(Fig~\ref{fig:zphot}), though we see no emission lines due to young
stars. At $z=$\zp$\sim$3.4 (Table~2), \hb\ and the \oiii\l\l4959,5007
doublet are expected to fall in the \k-band and Mg II \l2797 in the
\j-band. Note that the \k-band morphology clearly indicates an
extended source. At $z=$\zp, a half-light diameter $\approx$7 kpc is
implied, which is large; but such large values have been reported in the literature (e.g., Ivison
et al. 2001).

A\,2204\_2 is a soft source (Table 1). It is also at \zp=3.4, since
its colours and red $B-R$ limit are similar to A2204\_1. A single
\j-band spectrum only 1-hour long was obtained due to time-constraints
and no significant (non-spurious) features are found down to an EW$_{\rm
limit}^{3\sigma}\approx$ 17\AA\ (fourth spectrum from top in
Fig~\ref{fig:jspec}).


\begin{figure*}
  \begin{center}
 \includegraphics[angle=180,width=17.5cm]{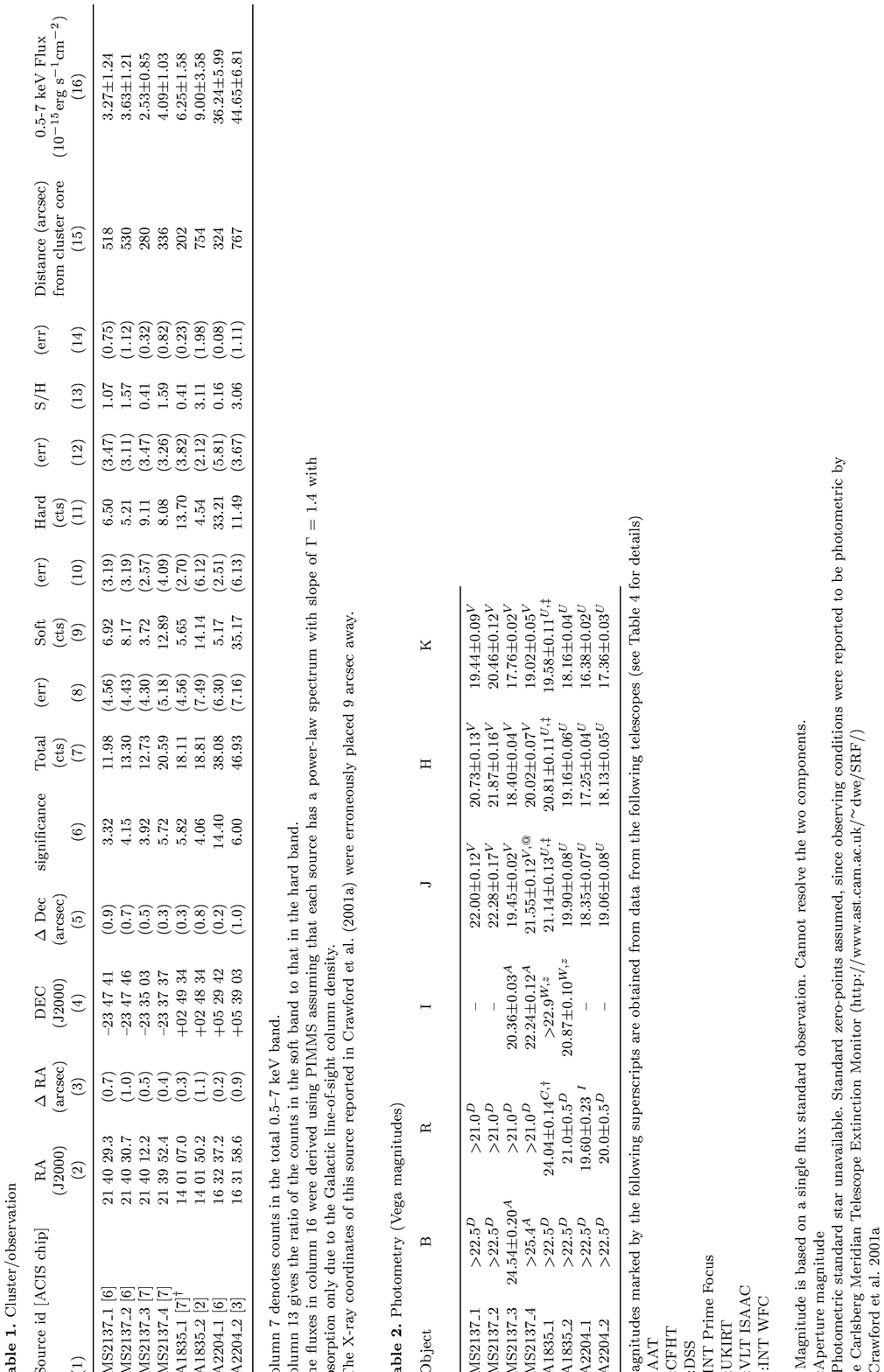}
  \label{fig:landphot}
  \end{center}
\end{figure*}

\addtocounter{table}{+2}
\onecolumn

\begin{table}
\caption{Near-IR Observation Log\label{tab:isaaclog}}
\begin{tabular}{lcccccccc}
Object    & RA         & DEC       & \multicolumn{3} {c} {Spectroscopic Observations} & \multicolumn{3} {c} {Imaging Exposures} \\
          & J2000      & J2000     & J  (s) &   K (s)   & Slit PA ($^{\circ}$)     & J  (s)   & H (s)     & K (s)\\
\hline 
MS2137\_1 & 21 40 29.3 & --23 47 39 & --     &  6120    & 111.0                    & 720      & 600       & 600\\
MS2137\_2 & 21 40 30.8 & --23 47 47 & --     &  6120    & 111.0                    & 720      & 600       & 600\\
MS2137\_3 & 21 40 12.2 & --23 35 03 & 7200   &  7200    & 113.5                    & 900      & 600       & 600\\
MS2137\_4 & 21 39 52.4 & --23 37 38 & 1800   &  7200    & 39.0                     & 720      & 600       & 600\\
A1835\_1  & 14 01 06.9 & +02 49 33  & --     &  7200    & 90.0                     & 540      & 540       & 540\\
A1835\_2  & 14 01 49.9 & +02 48 36  & --     &  5400    & 306.0                    & 540      & 540       & 540\\
A2204\_1  & 16 32 37.4 & +05 29 42  & 3780   &  7200    & 356.5                    & 540      & 540       & 540\\
A2204\_2  & 16 31 58.8 & +05 39 02  & 3600   &   --     & 47.0                     & 540      & 540       & 540\\
\hline
\end{tabular}
~~~\par
The coordinates denote the position measured in the near-IR pre-imaging, which was carried out at the VLT for objects in the field of MS2137, and at UKIRT for the rest of the sources.\\
The slit position angles are given in degrees east of North.
\end{table}

\begin{table}
\caption{Photometric Observations\label{tab:obsinfo}}
\begin{tabular}{lclclccc}
              && & & &  \\
Band & UT Date & Telescope \& Instrument & Plate Scale & Filters & Seeing & Typical Exposure & m$_{lim}$\\
     &      &                         & (arcsec/pixel) &      & (arcsec) & (seconds) & \\
\hline 
B & 1993 Aug 12 & AAT Prime Focus &  0.391 & KPNO 1      & 1.3 & 600 & 25.4\\
R & 1994 Jun 07 & INT Prime Focus &  0.590 & Kitt Peak 3 & 2.8 & 600 & 20.7\\
R & 1998 Feb 26 & CFHT STIS2      &  0.439 & CFHT \#4609 & 1.1 & 600 & 25.5\\
I & 1993 Aug 12 & AAT Prime Focus &  0.391 & KPNO 3      & 1.0 & 600 & 23.9\\
I & 2000 May 01 & INT WFC Prime   &  0.333 & Sloan i     & 2.0 & 1198& 22.9\\
J, H, K & 2000 Aug 10/11 &  UKIRT UFTI & 0.091 & J98, H98, K98 & 0.6 & 540 & 21.4, 20.7, 20.0\\
J, H, K & 2001 Jun 28 &  VLT ISAAC & 0.148 & J, H, Ks & 0.3 & 600 & 22.6, 21.6, 20.7\\
\hline 
\end{tabular}
~~~
The last column states the limiting magnitude as defined by 3 times the background sky RMS in a 3 arcsec aperture.\\
\end{table}

\begin{table}
\caption{Photometric Redshifts \label{tab:zphot}}
\begin{tabular}{lcccccccc}
Source & $z_{\rm photometric}$ & $\chi^2$ (filters)  & Galaxy Type & Age
& $A_{V}$ & M$_B$ & $z_{\rm secondary}$  \\
             &(90 percent interval)& & &(Gyr)& & & \\
\hline
MS2137\_3  & 0.70 (0.6, 0.9) & 2.5 (6) & Burst & 0.7 & 0.6 & -21.7 & 3.7\\
MS2137\_4  & 3.33 (3.1, 3.4) & 2.7 (5) & Burst & 0.3 & 0.0 & -24.6 & 4.1\\
A\,1835\_1 & 1.23$^\dag$ (1.0, 1.5) & 8.7 (5) & Burst & 0.4 & 0.2 & -21.5 & 4.6\\
A\,1835\_2 & 3.40 (3.0, 4.0) & 5.0 (6) & QSO   & --  & 0.6 & -25.2 & 3.6\\
A\,2204\_1 & 3.40 (3.3, 3.6) & 0.8 (5) & Burst & 0.1 & 0.0 & -27.3 & 0.5\\
A\,2204\_2 & 3.39 (2.9, 4.1) & 1.2 (5) & Burst & 0.1 & 0.0 & -26.4 & 0.5\\
\hline 
\end{tabular}
~~~\par
The 90 percent \zp\ confidence interval in parentheses in column 2 assumes $\Delta\chi^2=2.7$.\\
The \lq filters\rq\ parentheses in column 3 states the number of photometric data points used in the fit.\\
M$_B$ in column 8 is the absolute Vega Magnitude in the {\sl
B} Bessell filter.\\
The Burst template is a single burst of star-formation at zero age followed by passive evolution.\\
The QSO template is from Francis et al. (1991).\\
$^\dag$With the inclusion of the $R$-flux, HYPERZ prefers this solution to the one inferred by Crawford et al. (2001;  based on 4 filters). 
\end{table}

\begin{table}
\caption{Observed Spectroscopic Features\label{tab:lines}}
\begin{tabular}{lcccccrr}
Object      & Feature   & Width ($\sigma$) & FWHM & EW  &  Flux& \multicolumn{2} {c} {Notes}\\
            & \micron   & \AA           & km s$^{-1}$& \AA  & $10^{-17}$\ergpspsqcm &  & \\
\hline 
MS2137\_1   & \l2.0846  & 18.5 [17, 20] & U  & 240  & 12.8 [12, 14] & \ha\ \l6563 & $=>z=2.176$\\
            & \l2.0916  & 18.5$^{\dag}$ & U  & 134  & 7.2 [6, 8] & \nii\ \l6584 &\\
            & $\approx$\l2.1355  & 18.5$^{\dag}$  & U  & 134 & 3.9 [2, 6] & \sii\ \l\l6717 $+$ 6731 & \\
A\,1835\_1  & \l2.4452  & 18.1 [16, 20] & U  & 157 & 8.7 [8, 10]  & He I\l10830& $=>z=1.256$\\
            & \l2.2673  & 20.9 [11, 32] & U     & 32  & 1.9 [1, 3] & H I Pa 7 \l10049 &\\
A\,1835\_2  & \l2.3480  & 152.5 [134, 176] & 4580 & 207 & 36.9 [33, 41] & \hb\l4861 & $=>z=3.830$\\ 
\hline
\end{tabular}
~~~\par
\lq U\rq\ in column 3 implies unresolved.\\
$^{\dag}$Velocity width fixed to that of \ha.\\
The numbers in square brackets give the 90 percent confidence interval for a single parameter.\\
Strengths stated in row 3 are for the combined \sii\ doublet.\\
The line width and equivalent-width {\sl in the rest-frame} will be lower than the above values by a factor of $(1+z)$.
\end{table}

\begin{table}
\begin{center}
\caption{Detectability of emission lines at $z\ltsim 1$\label{tab:z1lines}}
\begin{tabular}{lcccr}
           &   &   &  &\\
Transition & J & K & {\sl I(transition)/I(\hb)} & Reference\\
\hline 
\ha\l6563          & $0.71<z<1.06$ &               & 2.87 & (1)\\
\nii\l6584         & $0.70<z<1.05$ &               & $\approx$2.8 & (2),(1)\\
\sii\l6731         & $0.66<z<1.01$ &               & $\approx$0.5 & Mrk 1073; (3)\\
\siii\l9069        & $0.24<z<0.49$ & $1.18<z<1.71$ & 0.1 & (4),(5)\\
\siii\l9531        & $0.18<z<0.42$ & $1.07<z<1.58$ & 0.25 & (5)\\
He I $\lambda10830$& $0.03<z<0.25$ & $0.83<z<1.27$ & 0.78 & (3)\\
\pab\l12822        & $z<0.05$      & $0.54<z<0.92$ & 0.17 & (1)\\
\paa\l18756        & --            & $z<0.31$      & 0.35 & (1)\\
\hline 
\end{tabular}
~~~\\
\end{center}
Column 4 denotes the intensity with respect to that of \hb. Any observed reddening has been corrected for. The last column gives the reference for the intensity calculation and/or measurement as follows:\\
(1) Case B photoionization prediction\\
(2) Storchi-Bergmann 1991\\
(3) Rudy et al. 1989\\
(4) Osterbrock \& Veilleux 1989\\
(5) Goodrich, Veilleux \& Hill 1994
\end{table}

\end{document}